\documentclass[12pt,preprint]{emulateapj}

\newcommand{\be}{\begin{equation}}
\newcommand{\ee}{\end{equation}}
\newcommand{\ba}{\begin{eqnarray}}
\newcommand{\ea}{\end{eqnarray}}

\begin{document}
\title{Cosmological constraints from line intensity mapping with interlopers}

\author{Yan Gong$^{1*}$, Xuelei Chen$^{2,3,4}$, and Asantha Cooray$^5$}

\affil{$^{1}$ Key Laboratory of Space Astronomy and Technology, National Astronomical Observatories,\\ Chinese Academy of Sciences, Beijing 100012, China}
\affil{$^{2}$ Key Laboratory of Computational Astrophysics, National Astronomical Observatories,\\ Chinese Academy of Sciences, Beijing 100012, China}
\affil{$^{3}$ School of Astronomy and Space Sciences, University of Chinese Academy of Sciences, Beijing 100049, China}
\affil{$^{4}$ Centre for High Energy Physics, Peking University, Beijing 100871, China}
\affil{$^{5}$Department of Physics and Astronomy, University of California, Irvine, CA 92697, USA}
\email{E-mail: gongyan@bao.ac.cn}

\begin{abstract}
Understanding the formation and evolution of the Universe is crucial for cosmological studies, and the line intensity mapping provides a powerful tool for this kind of study. We propose to make use of multipole moments of redshift-space line intensity power spectrum to constrain the cosmological and astrophysical parameters, such as the equation of state of dark energy, massive neutrinos, primordial non-Gaussianity, and star formation rate density. As an example, we generate mock data of multipole power spectra for H$\alpha\,6563\rm \AA$, [OIII]$\,5007\rm \AA$ and [OII]$\,3727\rm \AA$ measured by SPHEREx experiment at $z=1$ considering contaminations from interloper lines, and use Markov Chain Monte Carlo (MCMC) method to constrain the parameters in the model. We find a good fitting result of the parameters compared to their fiducial values, which means that the multipole power spectrum can effectively distinguish signal and interloper lines, and break the degeneracies between parameters, such as line mean intensity and bias. We also explore the cross power spectrum with CSST (Chinese Space Station Telescope) spectroscopic galaxy survey in the constraints. Since more accurate fitting results can be obtained by including measurements of the emission lines at higher redshifts out to $z=3$ at least and cross-correlations between emission lines can be involved, the line intensity mapping is expected to offer excellent results in future cosmological and astrophysical studies.

\end{abstract}

\keywords{cosmology: theory - large-scale structure of universe - cosmological parameters}

\maketitle

\section{Introduction}

A number of fundamental cosmological problems can be explored by galaxy surveys for probing cosmic large-scale structure. The ongoing and upcoming galaxy surveys, such as Sloan Sky Digital Survey (SDSS)\footnote{\tt https://www.sdss.org/}, The Dark Energy Spectroscopic Instrument (DESI)\footnote{\tt https://www.desi.lbl.gov/}, Large Synoptic Survey Telescope (LSST)\footnote{\tt https://www.lsst.org/} \citep{Ivezic08,Abell09}, {\it Euclid} space telescope\footnote{\tt https://www.euclid-ec.org/} \citep{Laureijs11} and Chinese Space Station Telescope (CSST) \citep{Zhan11, Zhan18, Cao18, Gong19}, will provide great information and insights on solving these problems. In traditional galaxy surveys, individual galaxies are resolved via high spatial resolution, and three dimensional (3D) or two dimensional (2D) angular correlation functions or power spectra in Fourier space can be derived for illustrating cosmic large-scale structure. However, these surveys are usually quite time-consuming to collect sufficient large galaxy sample, and especially, it is quite challenging for them to observe faint galaxies at high redshifts, which are precisely valuable and important for cosmological studies. By contrast, line intensity mapping provides a good option to overcome these difficulties.

Instead of observing individual galaxies, intensity mapping dedicates to measuring cumulative fluxes in a voxel defined by instrumental spatial and frequency resolutions. Therefore, fluxes no matter from  bright or faint galaxies in a voxel will be detected as signal in intensity mapping. Since huge amounts of galaxies can be included in a observed voxel, intensity mapping is quite efficient as a cosmological probe. Besides, because atomic and molecular emission lines are good tracers of galaxies, line intensity mapping is a suitable tool for measuring cosmic large-scale structure and galaxy formation and evolution. A number of works have discussed relevant issues about epoch of reionization (EoR) and post-EoR at $z<6$ \citep[e.g.][]{Visbal10,Carilli11,Gong11,Gong12,Gong13,Gong14,Gong17,Lidz11,Lidz16,Silva13,Silva15,Pullen14,Uzgil14,Yue15,Chen16,Fonseca16,Fonseca18,Padmanabhan18,Moradinezhad19}. However, there is a problem for line intensity mapping, that interloper lines redshifted into the same voxel of signal line can result in significant contaminations, and then it is difficult to distinguish signal line from interlopers.

A common method of reducing interloper contamination is cross-correlating intensity maps with other kinds of surveys, such as traditional galaxy surveys. Although this has been proved to be feasible \citep[e.g.][]{Chang10}, the auto correlation of signal line is hard to be directly measured in this method. Another way is masking the bright voxels in the survey volume, under the assumption that interloper lines are always much brighter than signal line. This method is simple and effective, but information in the masked voxels is wastefully discarded. On the other hand, if we have good understandings of interloper lines and could recognize specific features of them, they can be distinguished, and more importantly, can be seen as ``signals'' as well. That is to say, interloper lines potentially also can be used for extracting cosmological and astrophysical informations. \cite{Visbal10} and \cite{Gong14} find that the signal and interloper lines have different shapes in redshift-space line intensity power spectrum along wavenumbers perpendicular and parallel to the line of sight, which can be adopted for distinguish interlopers from signals. This method is further developed and discussed in details in \cite{Lidz16}. 

In this work, we explore the constraints on cosmological and astrophysical parameters using multipole moments of redshift-space line intensity power spectrum. As an example, we take multipole intensity power spectra of H$\alpha\, 6563\rm \AA$, [OIII]$\,5007\rm \AA$ and [OII]$\,3727\rm \AA$ measured by SPHEREx (Spectro-Photometer for the History of the universe, Epoch of Reionization, and Ices Explorer) experiment in the discussion. We consider time-variable equation of state of dark energy, massive neutrinos, and primordial non-Gaussianity in the cosmological model. We generate mock data of total multipole power spectra for the three emission lines with interlopers at $z=1$, and include the cross-correlation with CSST galaxy survey. The Markov Chain Monte Carlo (MCMC) method is adopted to constrain the parameters.

The paper is organized as follows: in Section~\ref{sec:CM}, we show the detailed cosmological models we consider in this study. In Section~\ref{sec:LIM}, we discuss the estimate of line mean intensity. In Section~\ref{sec:lps}, the calculations of multipole moments of intensity power spectrum of signal and interloper lines have been shown. In Section~\ref{sec:LD}, we generate mock data of multipole intensity power spectra based on measurements by SPHEREx experiment. In Section~\ref{sec:CC}, we discuss cross-correlation with CSST galaxy survey. In Section~\ref{sec:results}, we show the fitting results of cosmological and astrophysical parameters involved in the model. We summarize our results in Section~\ref{sec:summary}.

\section{Cosmological Model}
\label{sec:CM}

We assume a flat space of the Universe in this work, and consider dark energy model with time-variable equation of state, massive neutrinos, and primordial non-Gaussianity in the cosmological model. The details of the model are discussed as follow.

\subsection{Dark Energy}

The properties of dark energy can be represented by its equation of state $w=p/\rho$, where $p$ and $\rho$ are the pressure and energy density, respectively. The equation of state of dark energy can takes the values $w<-1$ (e.g. phantom), $w=-1$ (cosmological constant) and $-1<w<0$ (e.g. quintessence). In our model, we make use of a time-variable equation of state of dark energy, i.e. Chevallier-Polarski-Linder (CPL) parameterization \citep{Chevallier01, Linder03}, which takes the form as 
\be
w(z) = w_0 + \frac{w_a\,z}{1+z},
\ee
where $w_0$ and $w_a$ are the free parameters. As measured by current cosmological observations, $w_0$ and $w_a$ should be around -1 and 0, respectively. Then the Hubble parameter in the flat space can be calculated by
\ba
H(z) =&& H_0[\Omega_{\rm M}(1+z)^3+(1-\Omega_{\rm M}) \nonumber\\
          &&\times (1+z)^{3(1+w_0+w_a)}e^{-3w_a z/(1+z)}]^{1/2}.
\ea
Here $H_0=100\, h\ \rm km\,s^{-1} Mpc^{-1}$ is the Hubble constant. The Hubble parameter can characterize the kinetic expansion of the Universe. One the other hand, the dynamic evolution of the  structure of matter distribution can be evaluted by the linear growth factor for matter perturbation modes, which is given by \citep{Heath77,Peebles80}
\be
g(a) = \frac{5\,\Omega_{\rm M}}{2} \frac{H(a)}{a\,H_0} \int_0^a\frac{da'}{a'^3\,[H(a')/H_0]^3},
\ee
where $a=1/(1+z)$ is the scale factor. When calculating the matter power spectrum, the normalized growth factor at $z=0$ is always adopted, and it is defined as
\be
D(z) \equiv \frac{1}{1+z} \frac{g(z)}{g(0)}.
\ee
Finally, the linear matter power spectrum can be estimated as
\be
P^{\rm lin}_{\rm m}(k,z) = A_{\rm s} k^{n_{\rm s}} T^2(k) D^2(z),
\ee
where $A_{\rm s}$ is the primordial amplitude which can be replaced by the amplitude of current fluctuation on 8 Mpc$\,h^{-1}$ scale (i.e. $\sigma_8$), $n_{\rm s}$ is the primordial spectral index, $T(k)$ is the transfer function. As we see later, the linear matter power spectrum is suitable and good enough for our discussion, since we are mainly focusing on linear regime.

\subsection{Massive Neutrinos}

Neutrinos are relativistic and couple with other species in the early Universe when radiation is dominant. As the Universe expands and cools down, they decouple and redshift adiabatically. At that time, the relativistic neutrinos travel at the speed of light, but when they become non-relativistic, the thermal velocity decreases to
\be
v_{\rm th}(z) \simeq \frac{3T_{\nu}}{m_{\nu}} \simeq 151(1+z) \left( \frac{1\, {\rm eV}}{m_{\nu}} \right)\ \ {\rm km\,s^{-1}}.
\ee
Here $m_{\nu}$ is neutrino mass, and $T_{\nu}$ is neutrino temperature. As collisionless fluid,  the non-relativistic sub-eV neutrinos act as hot dark matter, that can free-stream from high to low matter density  regions and suppress fluctuations at scales smaller than the thermal free-streaming length. The wavenumber of free-streaming is given by
\be
k_{\rm FS}(z) = \sqrt{\frac{3}{2}} \frac{H(z)}{v_{\rm th}(z)(1+z)}.
\ee
Given low neutrino energy density, the suppressing of the matter power spectrum at small scales $k>k_{\rm FS}$ can be approximated as \citep{Hu98}
\be \label{eq:dP_nu}
\frac{\Delta P_{\rm m}}{P_{\rm m}} \simeq -8\frac{\Omega_{\nu}}{\Omega_{\rm M}},
\ee
where $\Omega_{\nu}=\sum m_{\nu}/(93.14\, h^2 {\rm eV})$ is the present neutrino energy density parameter. Note that the accurate suppressing fraction needs to be obtained by numerically solving the Boltzmann equation, and Eq.~(\ref{eq:dP_nu}) is only valid for small neutrino fraction with $f_{\nu}=\Omega_{\nu}/\Omega_{\rm M}\lesssim0.07$ (or $\sum m_{\nu}\lesssim1$ eV) \citep[see e.g.][]{Brandbyge08,Bird12}. For simplicity, we will adopt it in the discussion, since it is a good approximation as we show in Section~\ref{sec:results}, and should be sufficient for the purpose of this study.

For neutrinos becoming non-relativistic during matter domination era, the free-streaming scale leads to a maximum scale, whose wavenumber is given by
\be
k_{\rm nr} \simeq 0.018 \left( \frac{m_{\nu}}{1\,{\rm eV}}\right)^{1/2} \Omega_{\rm M}^{1/2}\ h\, {\rm Mpc^{-1}}.
\ee
On the scales much larger than the free-streaming scale, i.e. $k<k_{\rm nr}$, the neutrino thermal velocity is less than the escape velocity of gravitational potential wells, and does not affect matter fluctuations. This means that, on these scales, neutrino perturbations are identical to perturbations of cold dark matter.  Therefore, different neutrino mass can only significantly affect the matter power spectrum at small scales where $k>0.1$ Mpc$^{-1}$$h$ in practice. Then we can calculate the suppressed matter power spectrum with massive neutrinos by the formulae shown above.

\subsection{Primordial non-Gaussianity}

The primordial fluctuation is the seed of the cosmic large-scale structure. It is usually related to a inflation period in the very early Universe. The standard single-field slow-roll inflation model predicts that primordial fluctuations should be Gaussian distributed \citep{Maldacena03,Acquaviva03,Creminelli03}. However, other models such as multi-field inflation can result in significant primordial non-Gaussianity \citep{Linde97}. This leads the density fluctuations to be
\be
\Phi({\bf x}) = \phi({\bf x})+f_{\rm NL}\left[ \phi^2({\bf x})-\langle \phi^2\rangle \right],
\ee
where $\Phi({\bf x})$ is Bardeen's gauge-invariant potential at position $\bf x$, $\phi$ is Gaussian random field, and $f_{\rm NL}$ is the parameter indicating the overall amplitude of primordial non-Gaussianity.

The primordial non-Gaussianity can be described by high-order correlation functions, such as bispectrum in Fourier space. Generally speaking, it has three shapes, i.e. local, equilateral, and orthogonal. Here we will focus on the local shape, which has a distinct scale-dependent bias for the power spectra of tracers. This bias can be written as a linear bias with a scale-dependent correction, which is given by
\be
b^{\rm NG}(M,k,z) = b(M,z) + \Delta b(M,k,z).
\ee
The scale-dependent correction can be estimated by \citep{Dalal08,Slosar08}
\be \label{eq:b_NG}
\Delta b(M,k,z) = f_{\rm NL}[b(M,z)-1] \delta_{\rm c} \frac{3 \Omega_{\rm M}H_0^2}{k^2T(k)D(z)c^2},
\ee
where $\delta_{\rm c}=1.686$ is the density contrast factor for a spherical collapse of an overdensity region, $T(k)$ is the transfer function, $D(z)$ is the growth factor normalized at $z=0$, and $c$ is the speed of light. As we discuss in \S\ref{subsec:sps}, when using dark matter halos as tracers, $b^{\rm NG}(M,k,z)$ can be calculated by the halo model, and it will cause the bias of emission line to be scale-dependent in intensity mapping. Since the scale-dependent bias correction $\Delta b$ is only considerable at large scales, the primordial non-Gaussianity can be only effectively constrained at $k<0.02$ Mpc$^{-1}$$h$.

\section{line mean intensity}
\label{sec:LIM}

In this study, we consider four optical emission lines as signal and interloper lines, which are H$\alpha$\,6563$\rm \AA$, [OIII]\,5007$\rm \AA$, [OII]\,3727$\rm \AA$, and H$\beta$\,4861$\rm \AA$. As shown in \cite{Gong17}, the mean intensity of the lines can be estimated by three methods, i.e. observed line luminosity functions, cosmological simulations, and the star formation rate density (SFRD) derived from observations. These three methods are in good agreements in line intensity predictions, and we will adopt the SFRD method here since it is more convenient in our theoretical predictions. 

The line mean intensity as a function of redshift can be expressed as
\be \label{eq:I_SFR}
\bar{I}_{\rm line}(z) = \int_{M_{\rm min}}^{M_{\rm max}} {\rm d}M\frac{{\rm d}n}{{\rm d}M}\frac{L_{\rm line}(M,z)}{4\pi D_{\rm L}^2} y(z)D_{\rm A}^2,
\ee
where $M_{\rm min}=10^8$ ${\rm M}_{\sun}h^{-1}$ and $M_{\rm max}=10^{13}$ ${\rm M}_{\sun}h^{-1}$ are the minimum and maximum halo masses we use, ${\rm d}n/{\rm d}M\,(M,z)$ is the halo mass function \citep{Sheth99}, and $D_{\rm L}(z)$ and $D_{\rm A}(z)$ are the luminosity and comoving diameter distance at $z$, respectively. $y(z)={\rm d}r/{\rm d}\nu=\lambda_{\rm line}(1+z)^2/H(z)$, where $r$ is the comoving distance, $\lambda_{\rm line}$ is the rest-frame wavelength of emission lines, and $H(z)$ is the Hubble parameter. $L_{\rm line}(M,z)$ is the line luminosity, which can be related to the star formation rate (SFR). For the four emission lines we consider in this work, the $L_{\rm line}$-SFR relations are given by \citep{Kennicutt98, Ly07, Gong14, Gong17}
\ba 
{\rm SFR}\,(M_{\sun}{\rm yr^{-1}}) &=& (7.9\pm2.4)\times 10^{-42} L_{\rm H\alpha},\label{eq:SFR_Ha}\\
{\rm SFR}\,(M_{\sun}{\rm yr^{-1}}) &=& (7.6\pm3.7)\times 10^{-42} L_{\rm [OIII]},\label{eq:SFR_OIII}\\
{\rm SFR}\,(M_{\sun}{\rm yr^{-1}}) &=& (1.4\pm 0.4)\times 10^{-41} L_{\rm [OII]}.\label{eq:SFR_OII}
\ea
For H$\beta$ line, we adopt a relation H$\beta/$H$\alpha=0.35$ \citep{Osterbrock06}. This relation is found to be in a good agreement with observations and simulations \citep{Gong17}.

The SFR can be simply evaluated by assuming that it is proportional to halo mass $M$, which is a good approximation at $M\lesssim 10^{12}$ ${\rm M_{\sun}}$ \citep[see e.g.][]{Gong17}, and we have
\be \label{eq:SFR_M}
{\rm SFR}(M,z) = f_{\rm s}(z)\frac{\Omega_{\rm b}}{\Omega_{\rm M}}\frac{1}{t_{\rm s}}M,
\ee
where $t_{\rm s}=10^8$ yr is the typical star formation timescale, and $f_{\rm s}(z)$ is the the star formation efficiency at $z$, which can be estimated by
${\rm SFRD}(z) = \int {\rm d}M\frac{{\rm d}n}{{\rm d}M}{\rm SFR}(M,z)$.
Following \cite{Hopkins06}, we use the fitting formula \citep{Cole01}
\be \label{eq:SFRD_z}
{\rm SFRD}(z) = \frac{a+bz}{1+(z/c)^d}\,h\ (\,M_{\sun}{\rm yr^{-1} Mpc^{-3}}),
\ee
where $a=0.0118$, $b=0.08$, $c=3.3$ and $d=5.2$ with the initial mass function given by \cite{Baldry03}. 

Then we can calculate the line mean intensity using Eq.~(\ref{eq:I_SFR})-(\ref{eq:SFRD_z}). The uncertainties of the mean intensity are also considered by including the errors from the $L_{\rm line}$-SFR relations shown in Eq.~({\ref{eq:SFR_Ha}})-({\ref{eq:SFR_OII}}) and SFRD shown in Eq.~(\ref{eq:SFRD_z}) given by \cite{Hopkins06}. These uncertainties, including that of the relation between SFR and halo mass given by Eq.~(\ref{eq:SFR_M}), can affect the strength of line mean intensity, consequently change the amplitude of line intensity power spectrum, and finally impact the constraints on the cosmological and astrophysical parameters.

Besides, the dust extinction effect is also involved in this analysis. We make use of magnitude-averaged mean dust extinction laws, which give $A_{\rm H\alpha}=1.0$ mag, $A_{\rm [OIII]}=1.32$ mag, $A_{\rm [OII]}=0.62$ mag, and $A_{\rm H\beta}=1.38$ mag for the four lines we consider \citep{Kennicutt98,Calzetti00,Hayashi13,Khostovan15, Gong17}. The uncertainties and dust extinction effects of the line mean intensity will be passed into the estimates of line power spectra as shown in the next section.

\section{line intensity power spectrum}
\label{sec:lps}

In this section, we show the predictions of the signal power spectra of H$\alpha$, [OIII] and [OII] lines at $1\le z\le3$, and the observed power spectra of the three emission lines considering interlopers and uncertainties at $z=1$ as examples.

\subsection{Signal Power Spectrum}
\label{subsec:sps}

\begin{figure*}
\centerline{
\includegraphics[scale=0.27]{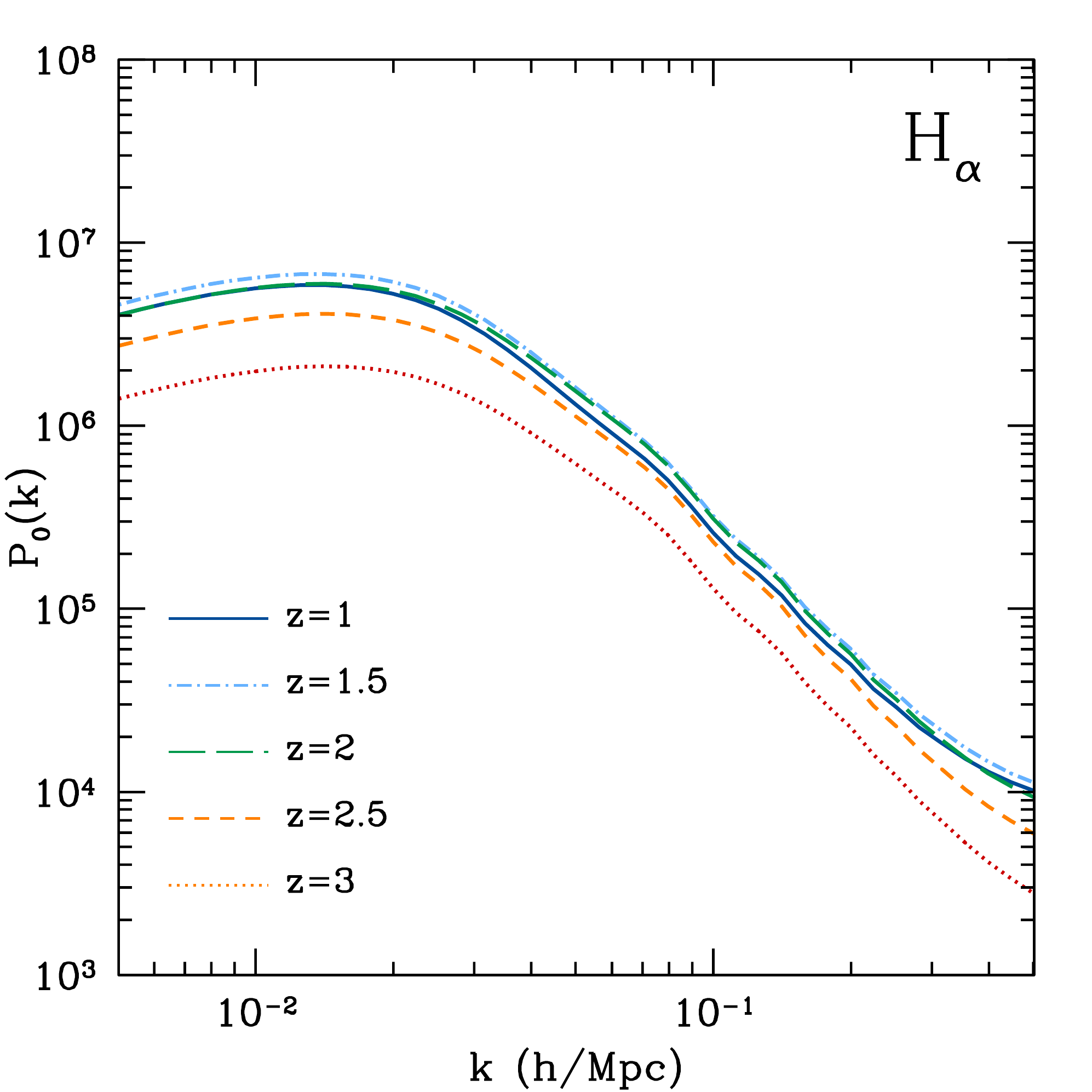}
\includegraphics[scale=0.27]{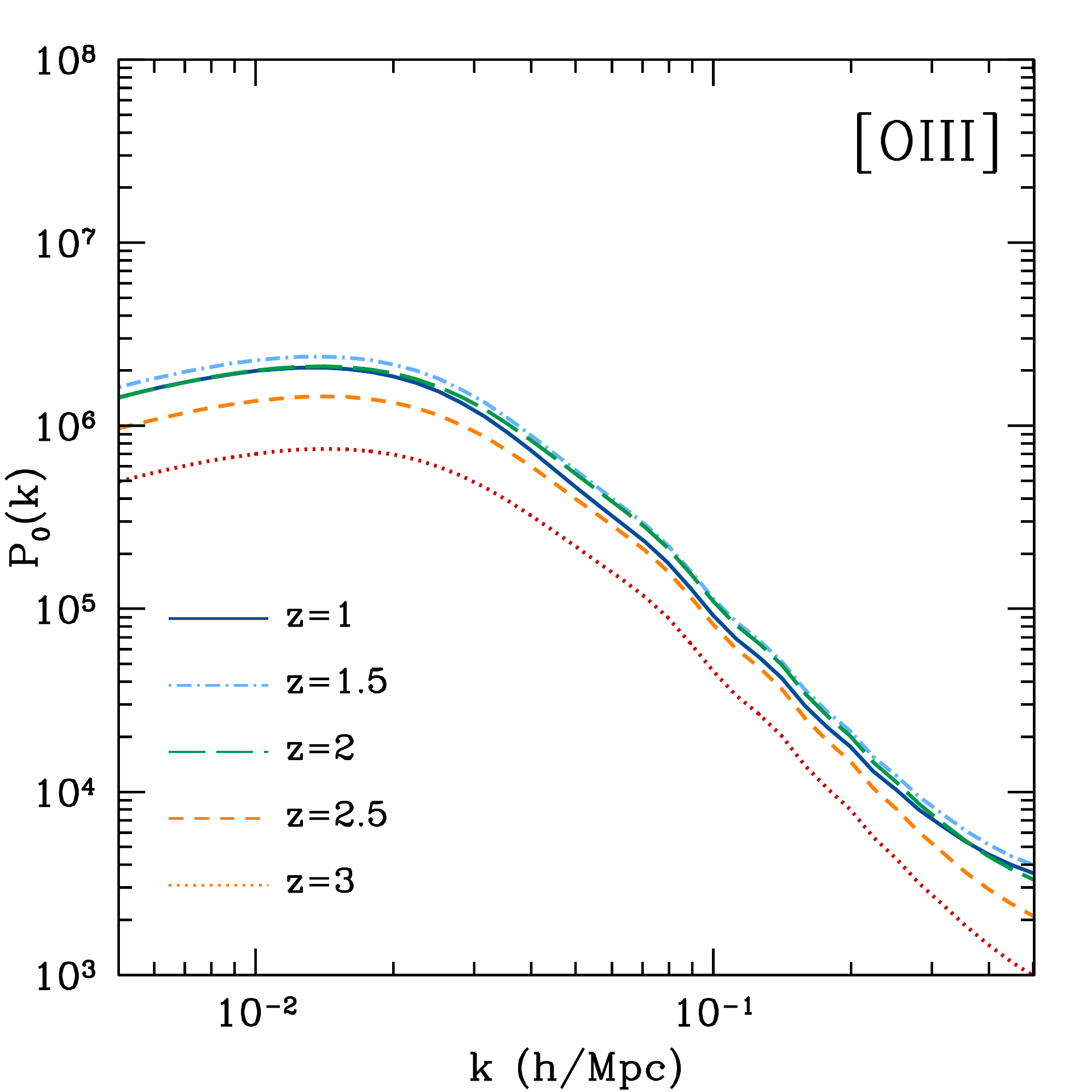}
\includegraphics[scale=0.27]{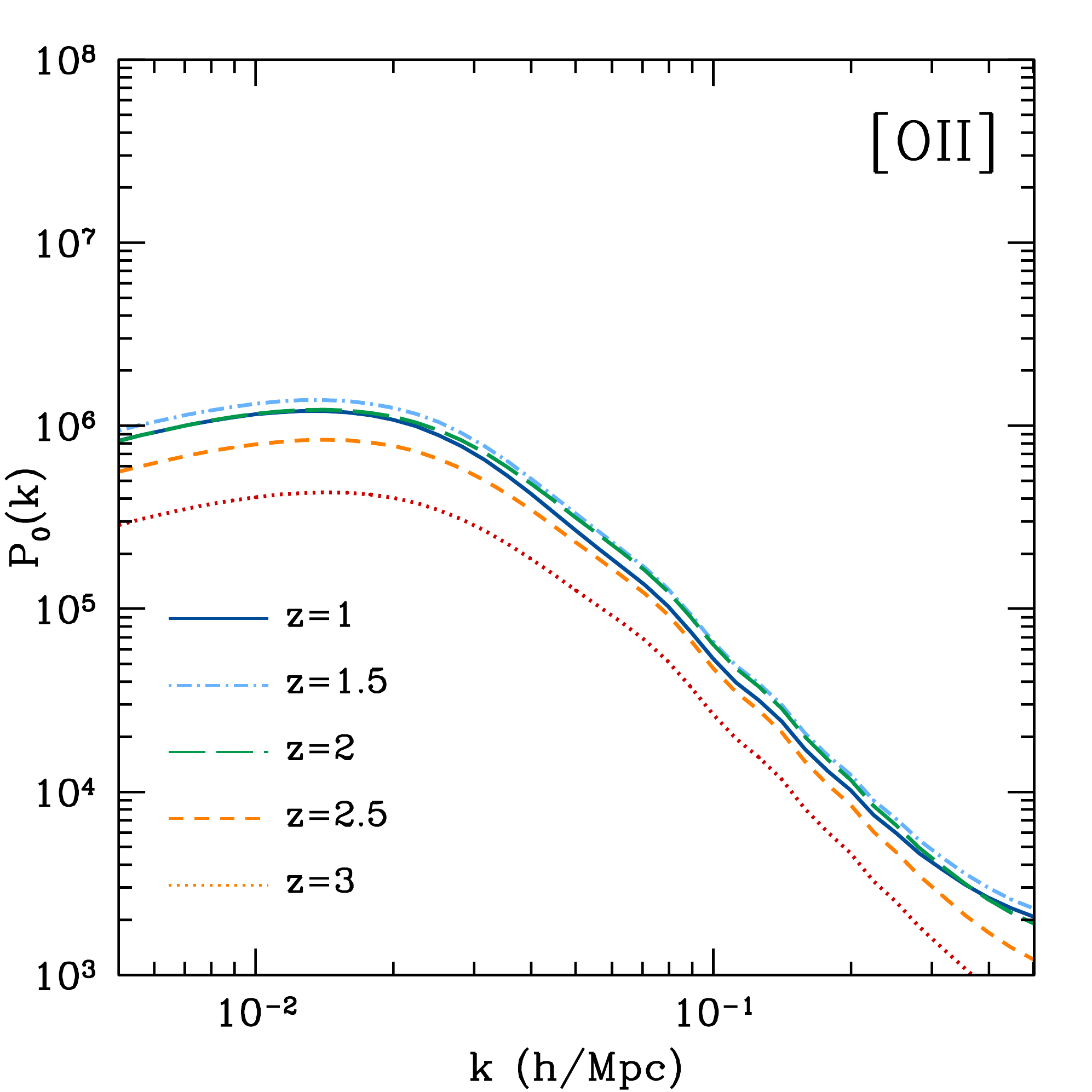}}
\centerline{
\includegraphics[scale=0.27]{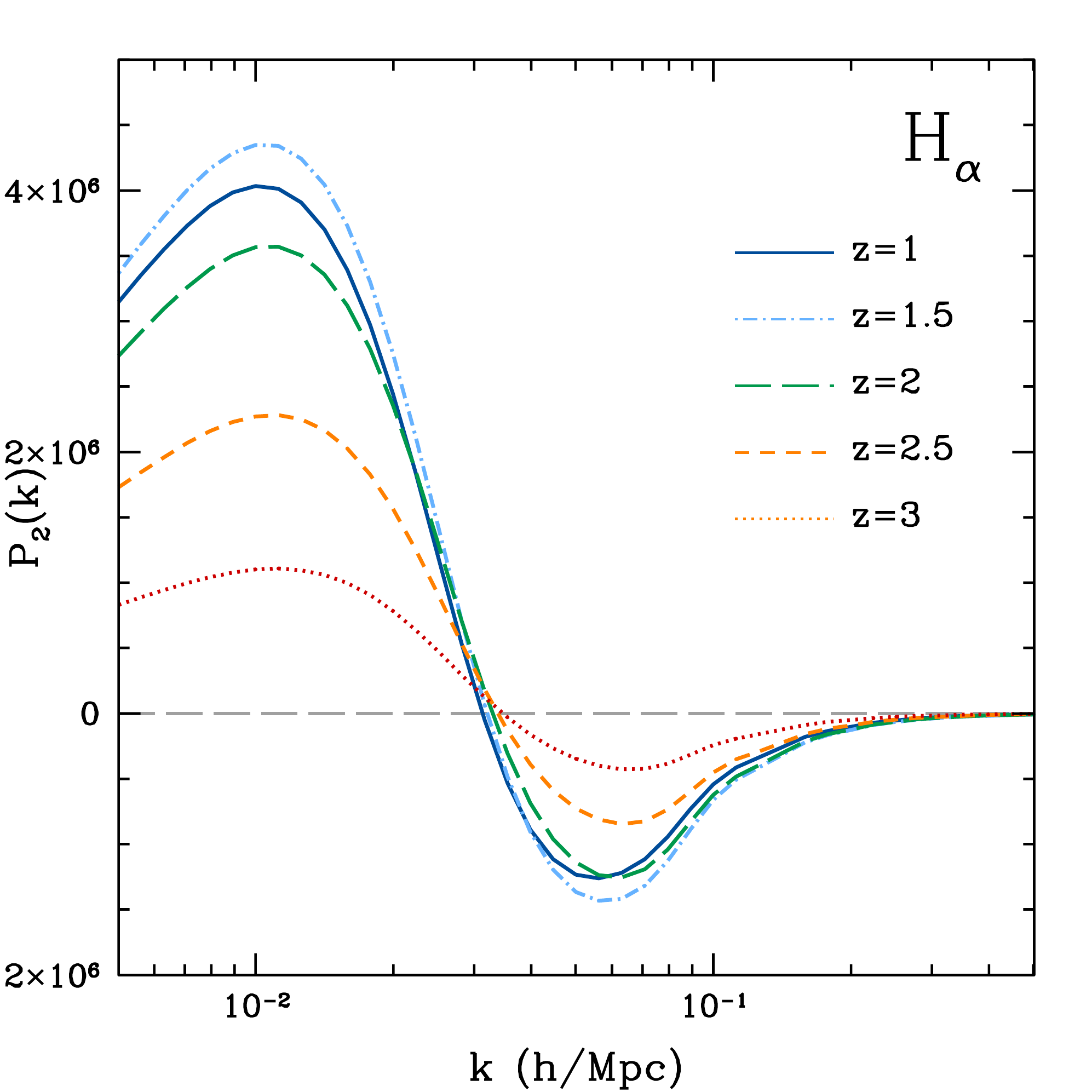}
\includegraphics[scale=0.27]{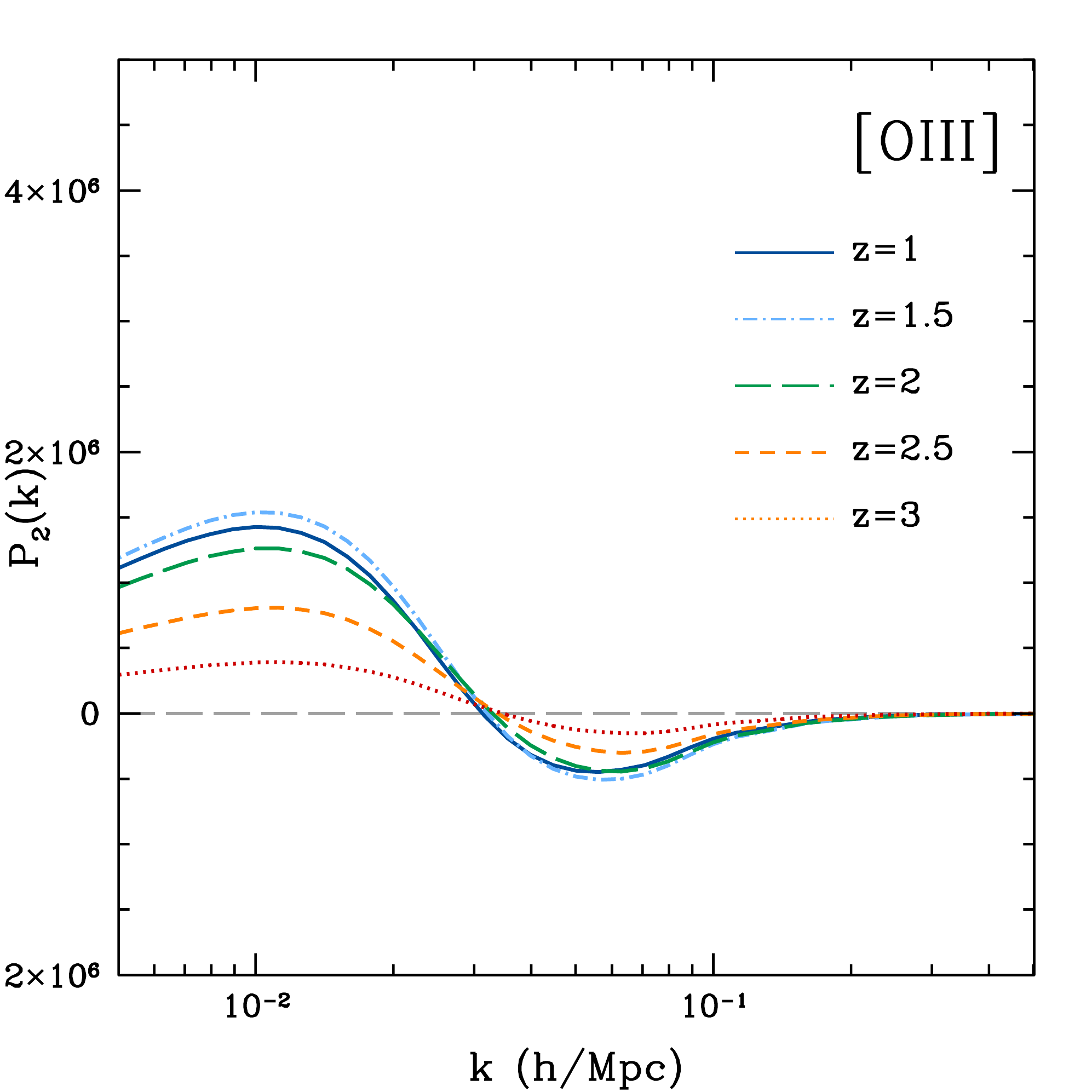}
\includegraphics[scale=0.27]{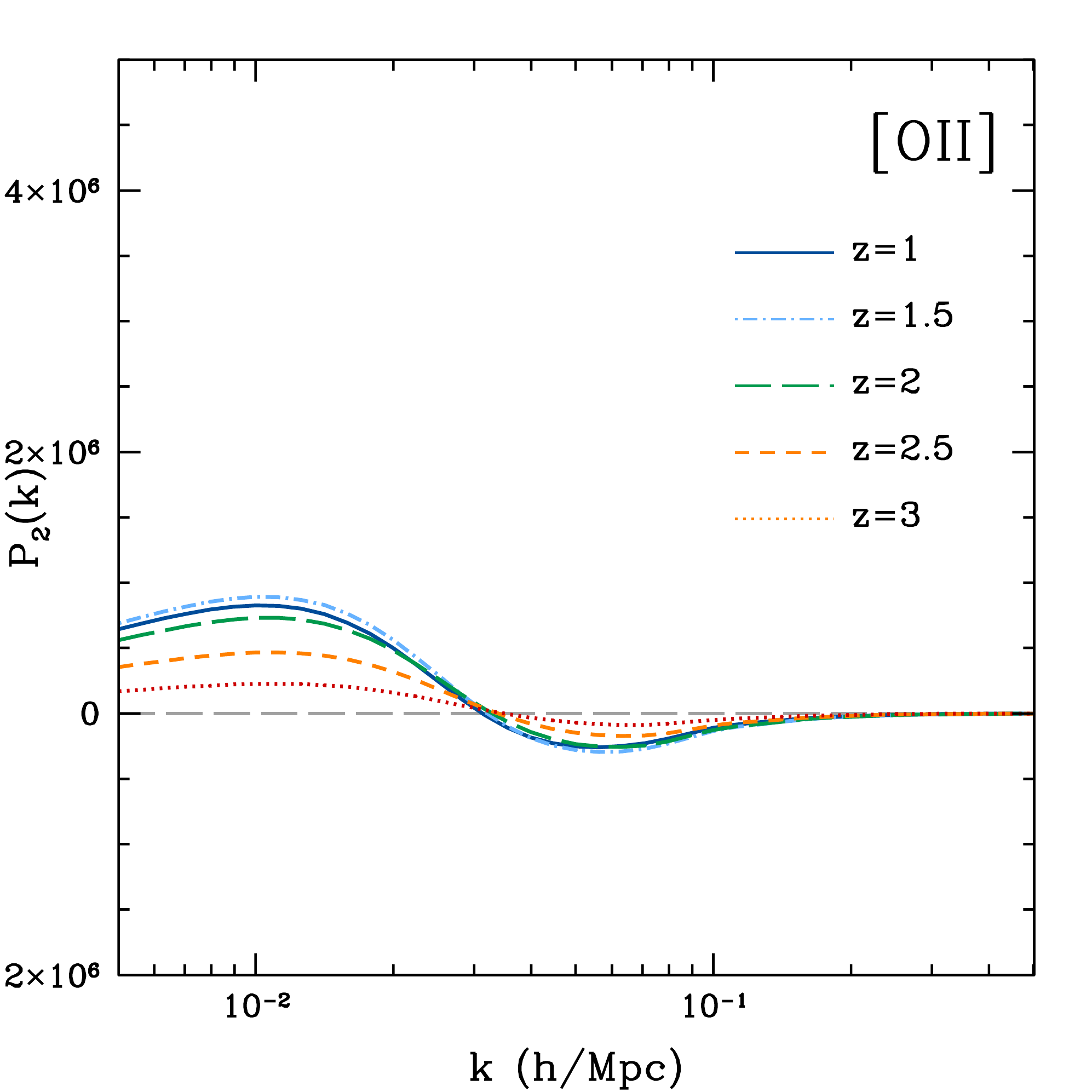}}
\centerline{
\includegraphics[scale=0.27]{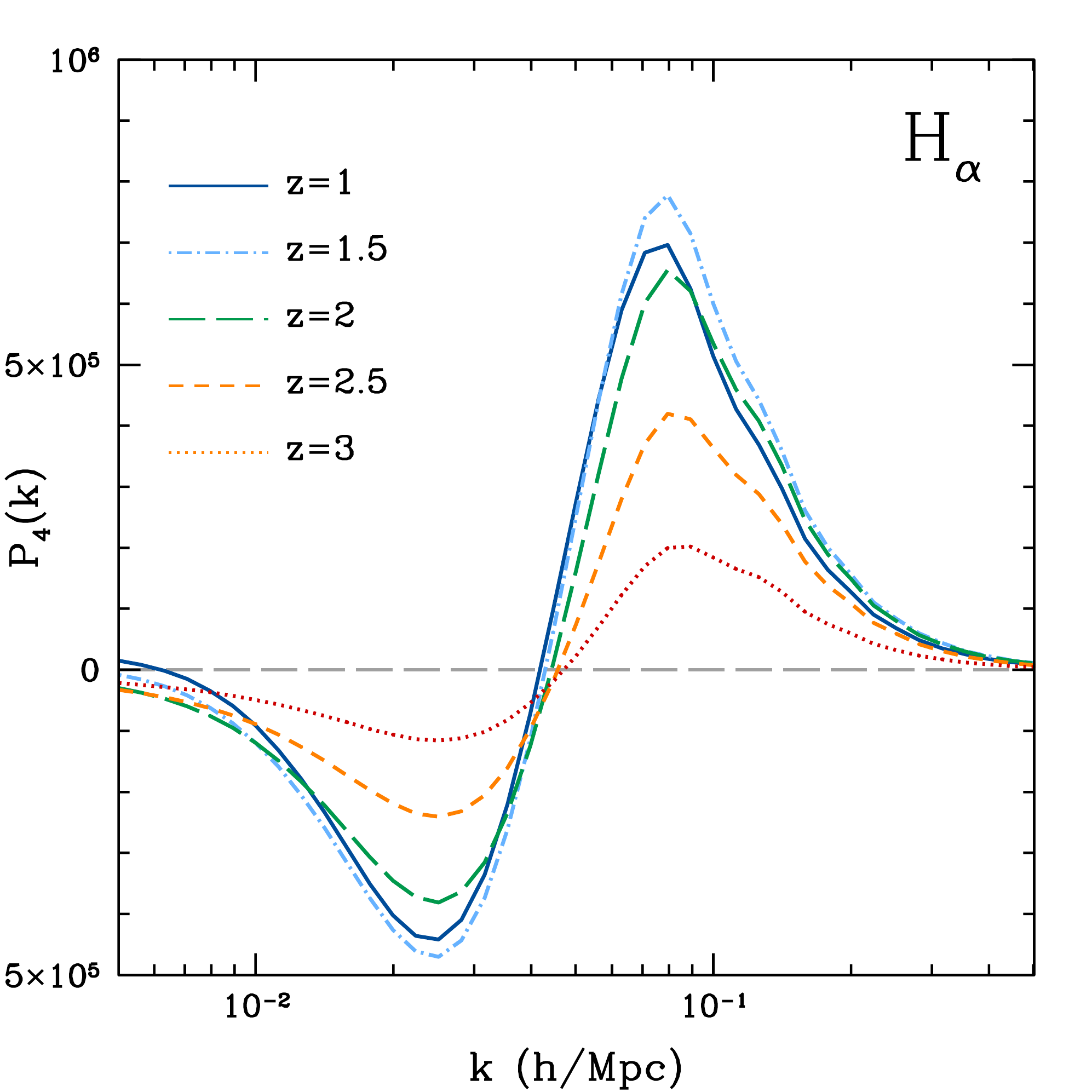}
\includegraphics[scale=0.27]{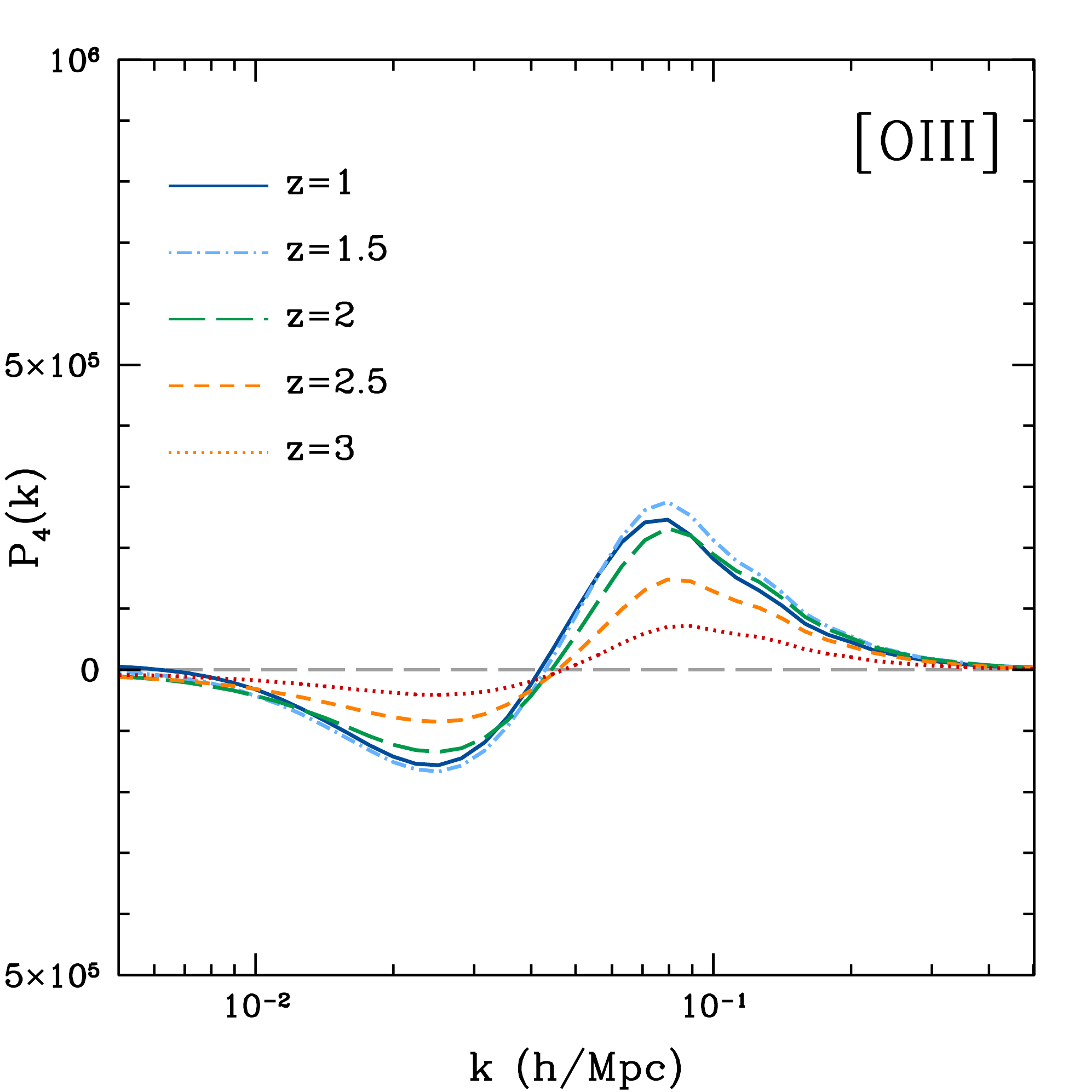}
\includegraphics[scale=0.27]{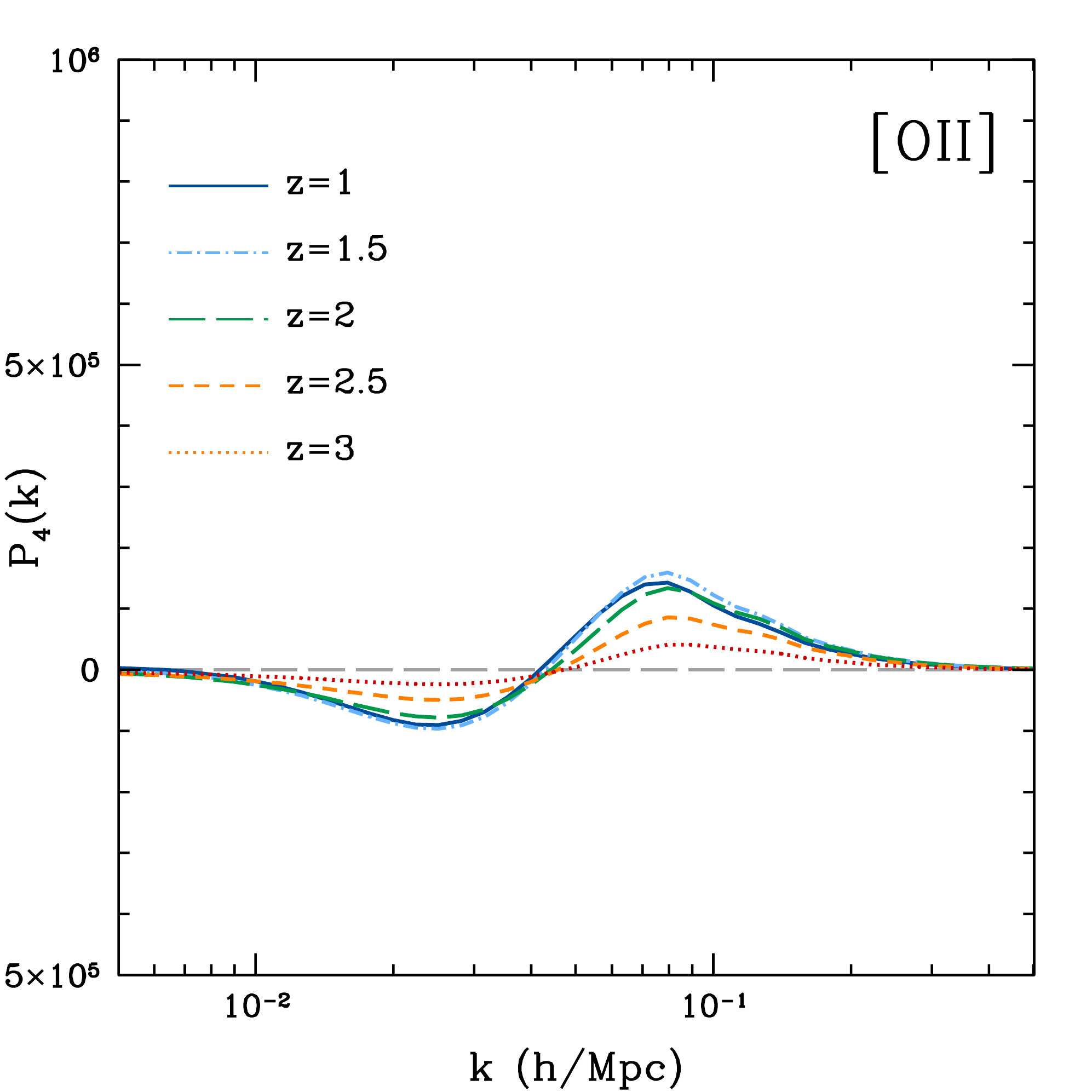}}
\caption{The multipole moments $P_0$ (top), $P_2$ (middle), and $P_4$ (bottom) of redshift-space power spectra of H$\alpha$ (left), [OIII] (middle), and [OII] (right) lines at $1\le z \le 3$ (in units of $\rm (Jy/sr)^2\,(Mpc/h)^3$). Note that $P_0$ is in logarithmic coordinates, while $P_2$ and $P_4$ are in linear coordinates to show the negative parts.}
\label{fig:Pl_lines_z}
\end{figure*}

We adopt multipole moments of redshift-space line intensity power spectrum as the estimator. Considering the Alcock-Paczynski effect (or AP effect) \citep{Alcock79}, it can be written as
\be \label{eq:Pl_line}
P^{\rm line}_{\ell}(k) = \frac{2\ell+1}{2\alpha^2_{\perp}\alpha_{\parallel}} \int_{-1}^1 {\rm d}\mu\, P^{\rm (s)}_{\rm line}(k',\mu') \mathcal{L}_{\ell}(\mu).
\ee
Here $\ell$ is the multipole, $k=\sqrt{k^2_{\parallel}+k^2_{\perp}}$ is the wavenumber, where $k_{\perp}$ and $k_{\parallel}$ are the components which are perpendicular and parallel to the line of sight, respectively. $\mu=k_{\parallel}/k$ is the cosine of the angle between the direction of wavenumber and the line of sight. $k'=\sqrt{k'^2_{\parallel}+k'^2_{\perp}}$ and $\mu'=k'_{\parallel}/k'$ are the apparent wavenumber and cosine of angle, where $k'_{\parallel}=k_{\parallel}/\alpha_{\parallel}$ and $k'_{\perp}=k_{\perp}/\alpha_{\perp}$. $\alpha_{\perp}=D_{\rm A}(z)/D^{\rm fid}_{\rm A}(z)$ and $\alpha_{\parallel}=H^{\rm fid}(z)/H(z)$ are the scaling factors in the transverse and radial directions, respectively. $D_{\rm A}(z)$ and $H(z)$ are the angular diameter distance and Hubble parameter at redshift $z$, respectively, and the superscript ``fid" means the quantities in the fiducial cosmology. $\mathcal{L}_{\ell}(\mu)$ is the Legendre polynomials that only the first three non-vanishing orders $\ell=(0, 2, 4)$ are considered here, and they take the values as 1, $1/2\,(3\mu^2-1)$, and $1/8\,(35\mu^4-30\mu^2+3)$, respectively. $P^{\rm (s)}_{\rm line}(k',\mu')$ is the apparent redshift-space line intensity power spectrum. By assuming that there is no peculiar velocity bias, it can be estimated by
\ba
P^{\rm (s)}_{\rm line}(k',\mu',z) &=& P_{\rm line}^{\rm clus}(k',z) (1+\beta\mu'^2)^2 \nonumber \\
                                                 &\times& \mathcal{D}(k',\mu',z) + P^{\rm shot}_{\rm line}(z),
\ea
where the superscript (s) denotes the quantity in redshift space. $P_{\rm line}^{\rm clus}(k',z)$ is the apparent real-space clustering line intensity power spectrum, and $P^{\rm shot}_{\rm line}(z)$ is the shot-noise power spectrum, which is not affected by the redshift-space distortion effect. $\beta=f/\bar{b}_{\rm line}(z)$ where $f={\rm d\,ln}D(a)/{\rm d\, ln}\,a$ is the growth rate. Here $D(a)$ is the growth factor normalized at $z=0$, and $\bar{b}_{\rm line}$ is the line mean bias. Note that this redshift-distortion effect also can help to break the degeneracy between the line bias and mean intensity \citep{Lidz16,Chen16}. The factor $\mathcal{D}(k', \mu')$ is the damping term at small scales, which is given by
\be
\mathcal{D}(k',\mu') = {\rm exp}\left[ -\left(k'\mu'\sigma_{\rm D}\right)^2\right].
\ee
Here $\sigma_{\rm D}$ denotes the effects of velocity dispersion and spectral resolution. In the linear regime at large scales where the intensity mapping focuses on, we find that this damping term is actually not important to affect the result. 

The clustering line intensity power spectrum,$P^{\rm clus}_{\rm line} $, which can be calculated by
\be
P^{\rm clus}_{\rm line}(k,z) = \bar{b}^2_{\rm line}(z) \bar{I}^2_{\rm line}(z) P_{\rm m}(k,z).
\ee
where $P_{\rm m}(k,z)$ is the matter power spectrum. Note that the matter power spectrum needs to be multiplied by a factor of $(1+\Delta P_{\rm m}/P_{\rm m})$ as indicated in Eq.~(\ref{eq:dP_nu}) when massive neutrinos are involved in the model. The mean line bias takes the form as
\be \label{eq:b_line}
\bar{b}_{\rm line}(z)=\frac{\int^{M_{\rm max}}_{M_{\rm min}} {\rm d}M \frac{{\rm d}n}{{\rm d}M}\, L_{\rm line}\, b(M,z)}{\int^{M_{\rm max}}_{M_{\rm min}} {\rm d}M \frac{{\rm d}n}{{\rm d}M} \,L_{\rm line}},
\ee
where $b(M,z)$ is the halo bias \citep{Sheth99}. When considering primordial non-Gaussianity, $b(M,z)$ should be replaced by $b^{\rm NG}(M,k,z)$ given by Eq.~(\ref{eq:b_NG}), and the mean line bias becomes scale-dependent as $\bar{b}_{\rm line}(z)\to \bar{b}_{\rm line}(k,z)$. The line shot-noise power spectrum is
\be\label{eq:Pshot_line}
P^{\rm shot}_{\rm line}(z) = \int_{M_{\rm min}}^{M_{\rm max}} {\rm d}M \frac{{\rm d}n}{{\rm d}M} \left[\frac{L_{\rm line}}{4\pi D_{\rm L}^2}y(z)D_{\rm A}^2\right]^2.
\ee

We consider H$\alpha$, [OIII], and [OII] as the signal lines in this study, since they are usually bright and relatively easy to be detected \citep{Gong17}. In Figure~{\ref{fig:Pl_lines_z}}, we show the multipole moments $P_0$, $P_2$, and $P_4$ of redshift-space power spectra of H$\alpha$, [OIII], and [OII] lines at $z=1$, 1.5, 2, 2.5, and 3. We find that the multipole power spectra at $z=1$, 1.5, and 2 has similar amplitude, while the ones at $z=2.5$ and $3$ declines significantly, which is due to the cosmic star formation history as indicated by SFRD$(z)$. In the following discussion, as examples, we will focus on the power spectra at $z=1$ to show the contaminations of interlopers, uncertainties, and detectability. At this redshift, dark energy begins to dominate the evolution of the Universe, the signals of line intensity mapping are relatively strong as shown in Figure~\ref{fig:Pl_lines_z}, and several spectroscopic galaxy surveys can be used to perform cross-correlation for further improving the strength of cosmological constraint as discussed in Section \ref{sec:CC}.

\subsection{Observed Power Spectrum}
\label{subsec:ops}

\begin{figure*}
\centerline{
\includegraphics[scale=0.27]{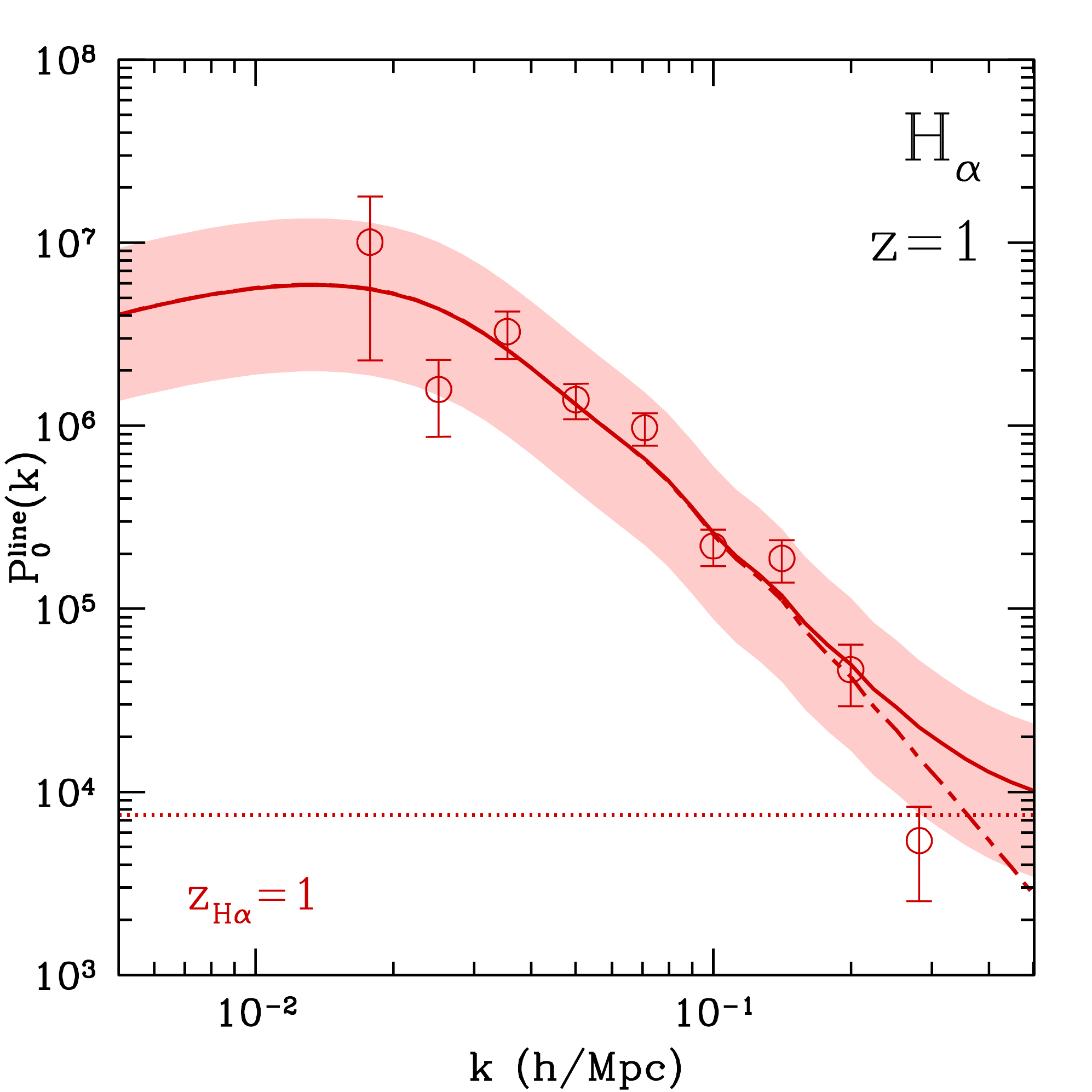}
\includegraphics[scale=0.27]{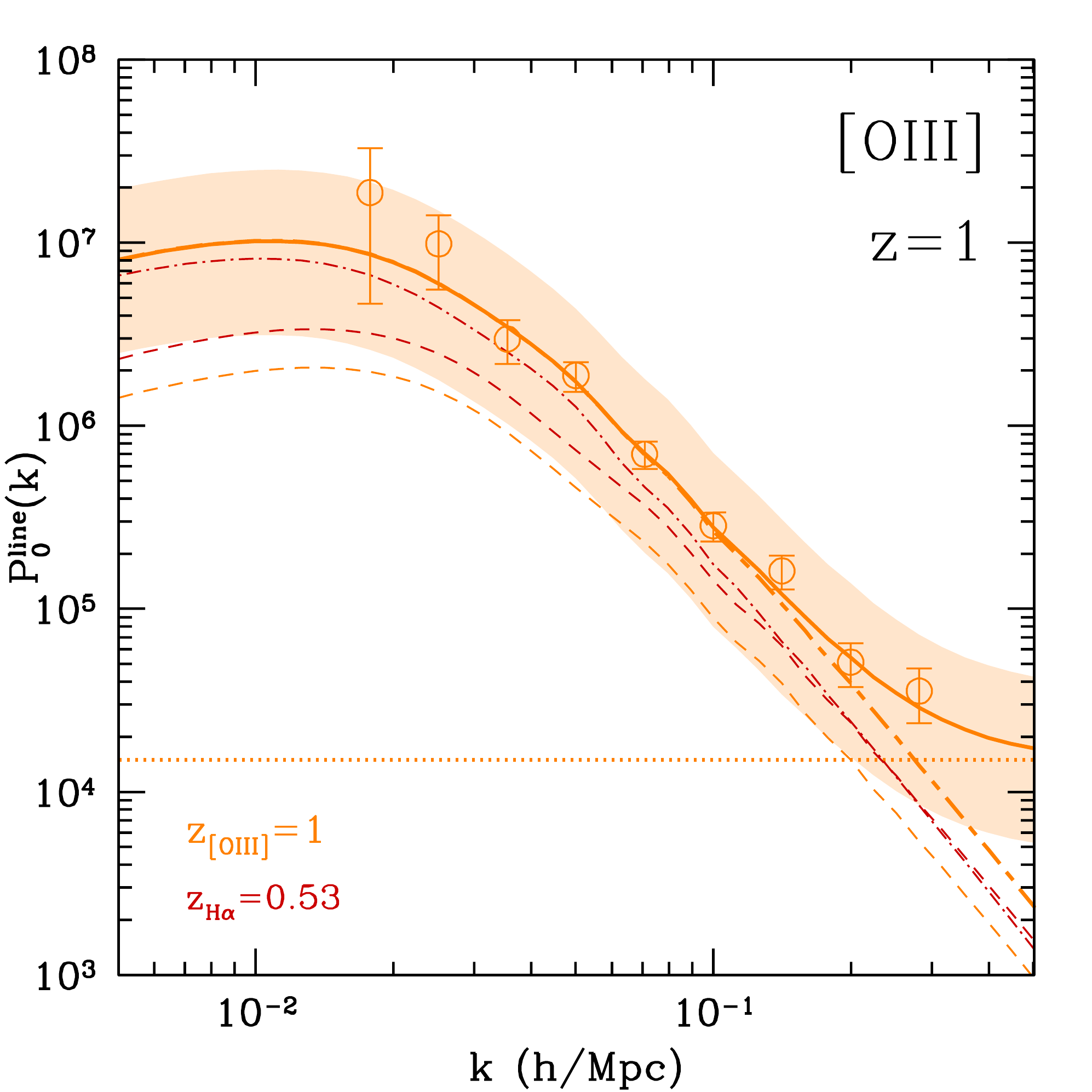}
\includegraphics[scale=0.27]{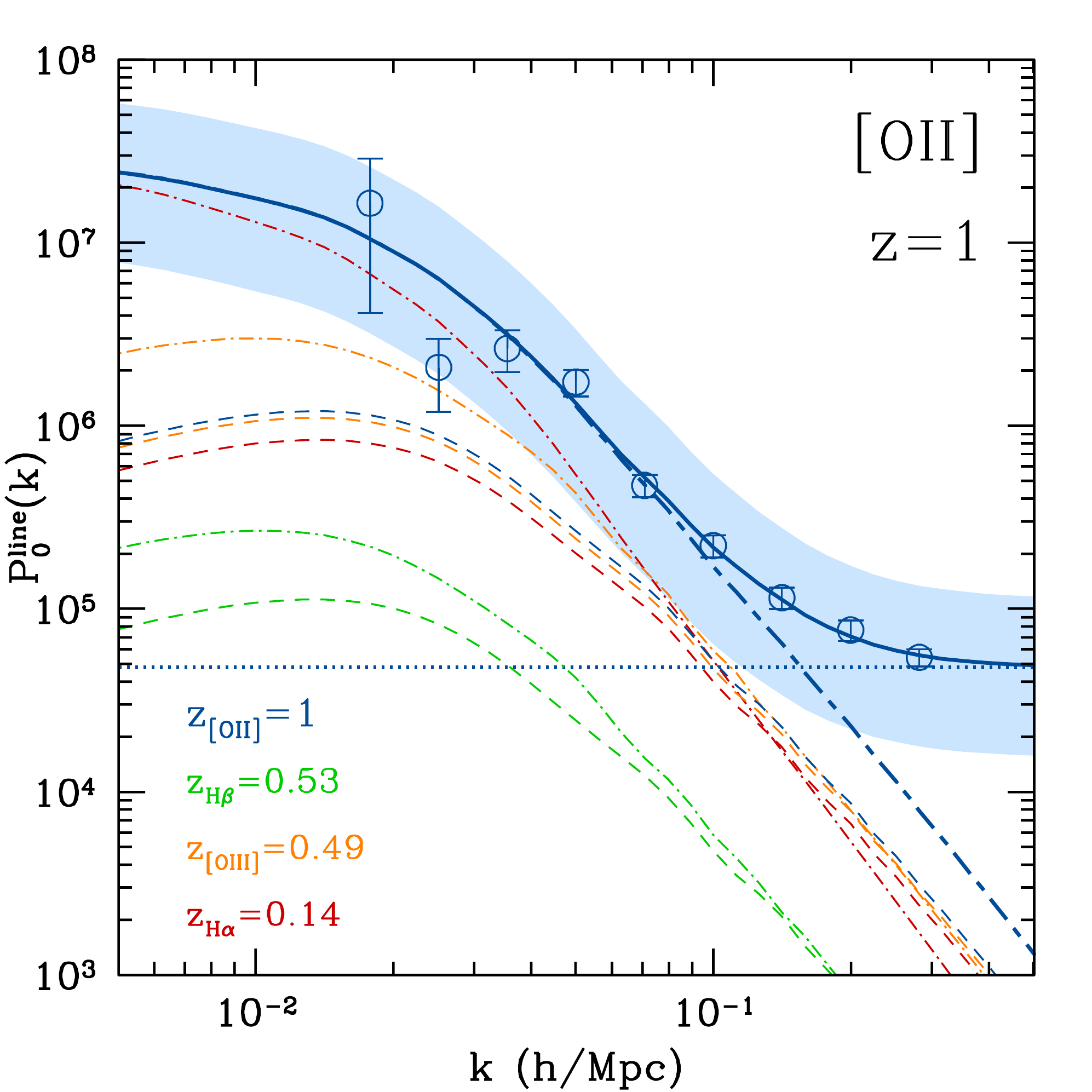}}
\centerline{
\includegraphics[scale=0.27]{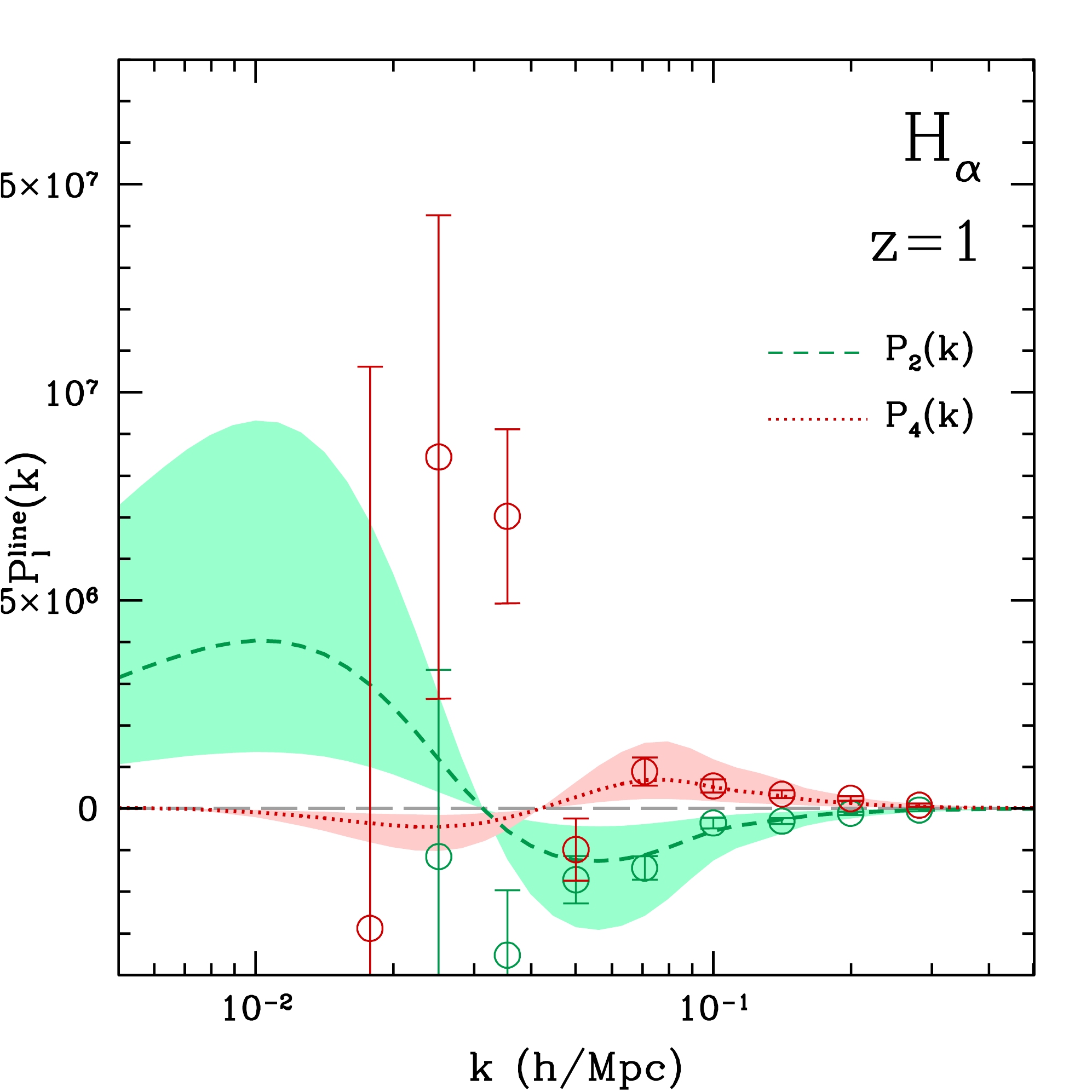}
\includegraphics[scale=0.27]{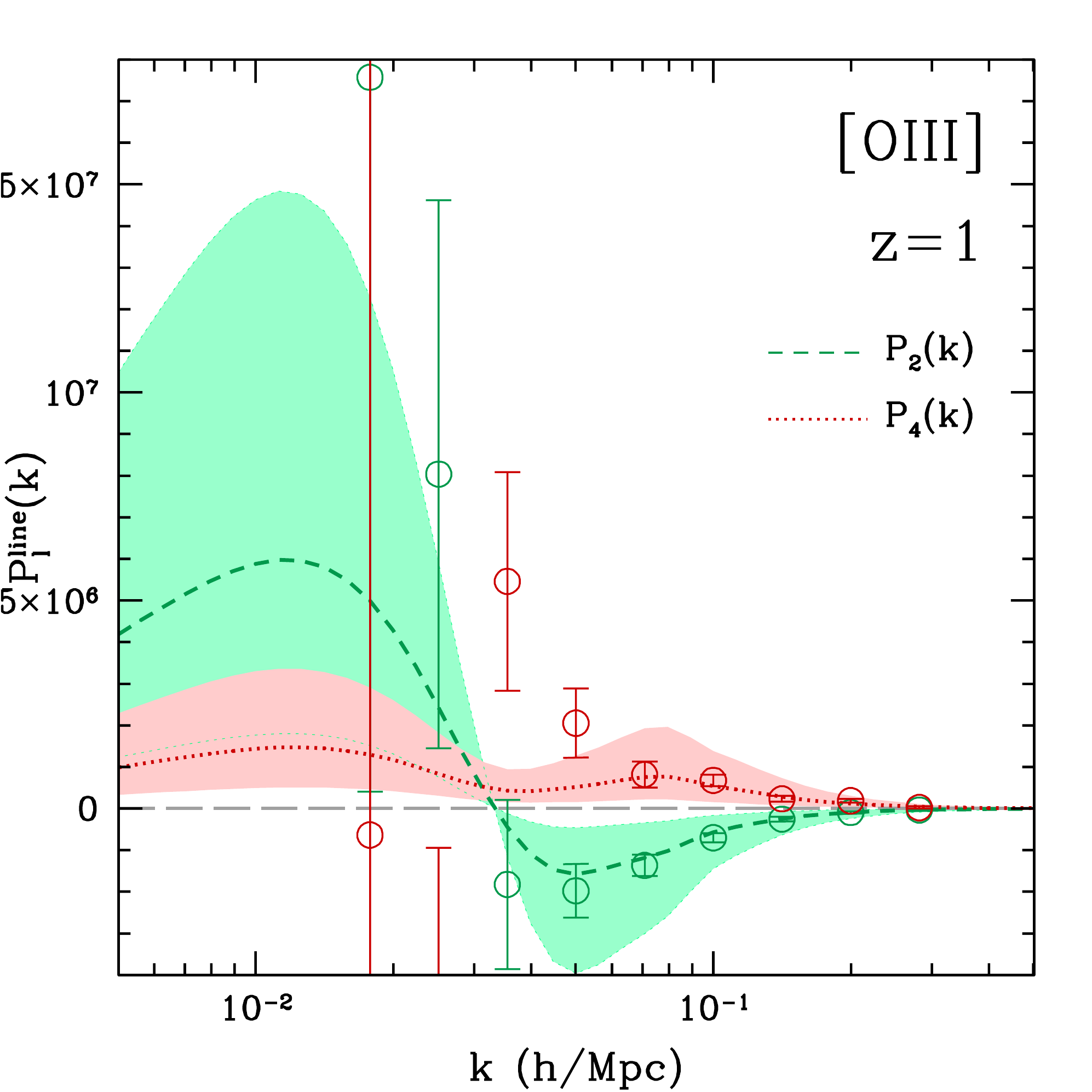}
\includegraphics[scale=0.27]{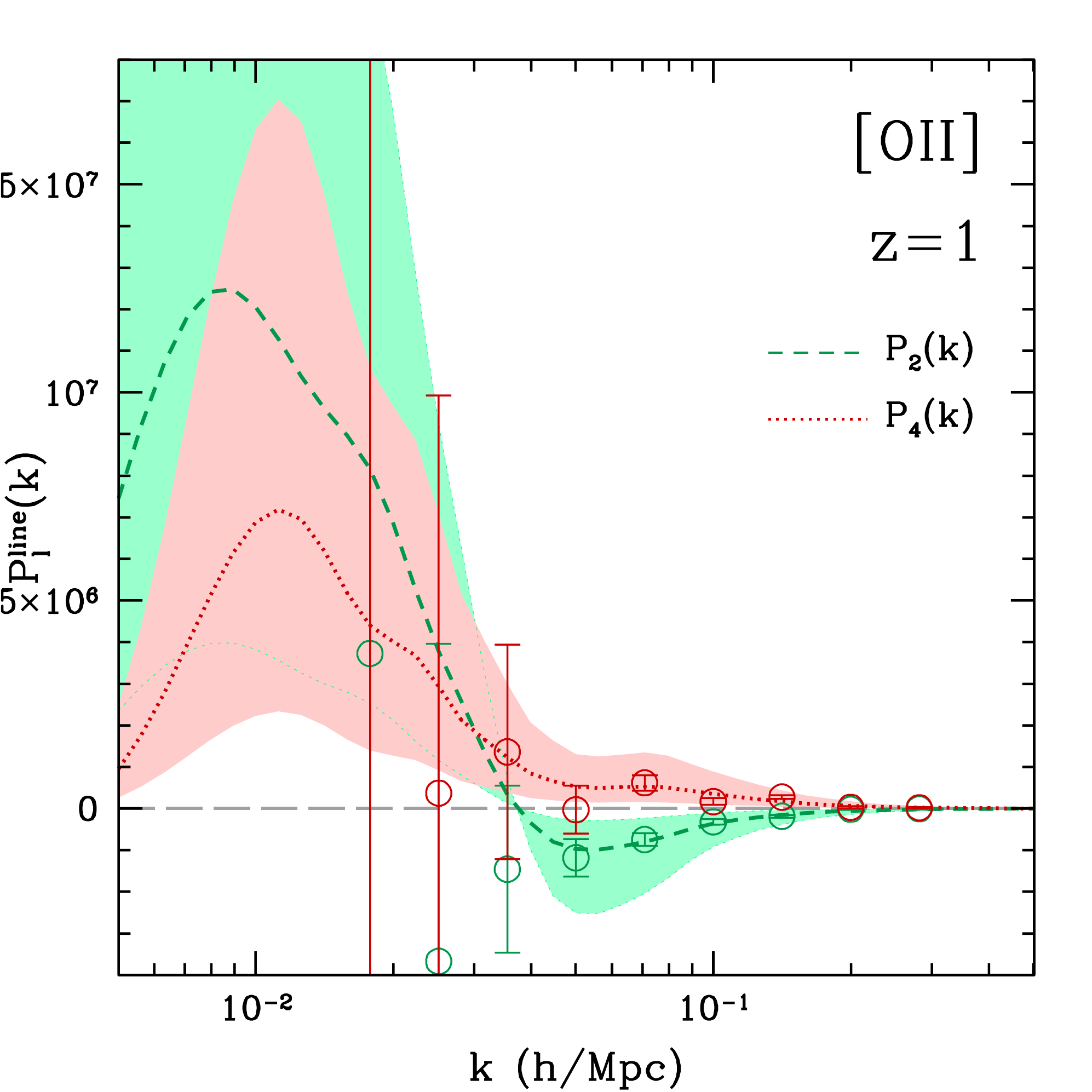}}
\caption{The observed multipole power spectra of H$\alpha$, [OIII] and [OII] at $z=1$ with interloper lines (in units of $\rm (Jy/sr)^2\,(Mpc/h)^3$). The shaded regions denote the uncertainties for the observed signal power spectra. The data points with error bars are estimated based on the SPHEREx experiment. $Top$: $P_0$ power spectra for the signal and interloper lines. The observed total power spectra (clustering+shot-noise), total clustering power spectra (signal+interlopers), and total shot-noise power spectra (signal+interlopers) at $z=1$ are shown as solid, long-short dashed, dotted curves, respectively. The red, orange and blue curves are for H$\alpha$, [OIII] and [OII] lines, respectively. The signal clustering power spectra at $z=1$ are shown in dashed curves. As comparison, the projected and original power spectra of interloper lines are shown in dash-dotted and dashed curves, respectively. $Bottom$: The observed total $P_2$ (green dashed) and $P_4$ (red dotted) power spectra and mock data points.}
\label{fig:Pl_err}
\end{figure*}

In observations, the signal of emission line can be contaminated by the continuum emission and interloper lines redshifted to the same frequency. The continuum contamination can be effectively removed by the smooth feature of its spectrum as a function of frequency \citep[e.g.][]{Silva15,Yue15}. Therefore, the main contaminations actually come from interloper lines, especially the ones at lower redshifts\footnote{As we show later, the contaminations of other lines from higher redshifts can be safely ignored for the three emission lines we study, especially considering the projection effect for interlopers.}. As discussed in \cite{Gong17}, the contamination on H$\alpha\,\rm 6563\AA$ can be neglected, H$\alpha\,\rm 6563\AA$ at lower redshift can contaminate [OIII]$\,\rm 5007\AA$, and H$\alpha\,\rm 6563\AA$, [OIII]$\,\rm 5007\AA$ and H$\beta\,\rm 4861\AA$ can be significant foregrounds for [OII]$\,\rm 3727\AA$ at higher redshift.

Considering the contaminations from interloper lines, the observed power spectrum of a emission line is composed of signal power spectrum and all components from interlopers, which is given by \citep[e.g.][]{Visbal10, Gong14, Lidz16, Gong17}
\be \label{eq:P_obs}
P_{\ell,\rm obs}(k,z) = P_{\ell,\rm s}(k,z) + \sum_{i=1}^N P_{\ell,\rm i}^{{\rm pro},i}(k_{\rm i},z).
\ee
Here $P_{\ell,\rm s}(k,z)$ is the signal power spectrum given by Eq.~(\ref{eq:Pl_line}). $P_{\ell,\rm i}^{{\rm pro},i}(k_{\rm i},z)$ is the $i$th interloper power spectrum, which is projected to the signal redshift $z$. $k_{\rm i}$ is the wavenumber at the redshift of a interloper line $z_{\rm i}$, and we have  $k_{\rm i}=\sqrt{A_{\perp}^2k_{\perp}^2+A_{\parallel}^2k_{\parallel}^2}$. $A_{\perp}$ and $A_{\parallel}$ are the factors to transfer $k$ to $k_{\rm i}$, which are given by $A_{\perp}=r_{\rm s}/r_{\rm i}$ and $A_{\parallel}=y_{\rm s}/y_{\rm i}$, where the subscripts ``s'' and ``i'' denote ``signal" and ``interloper", respectively. Then the projected interloper power spectrum can be calculated by projecting the interloper power spectrum at $z_{\rm i}$ to the signal redshift $z$, which is given by \citep{Visbal10, Gong14}
\be
P^{\rm pro}_{\ell,\rm i}(k_{\rm i},z) = A_{\perp}^2A_{\parallel} P_{\ell,\rm i}(k_{\rm i},z_{\rm i}).
\ee
Unlike the signal power spectrum, the projected interloper power spectrum $P^{\rm pro}_{\ell,\rm i}(k_{\rm i},z)$ is anisotropic in the $k_{\perp}-k_{\perp}$ space, which can be used to recognize and remove the effect of interloper lines \citep{Gong14,Lidz16}. 

In Figure~{\ref{fig:Pl_err}}, the observed multipole moments of power spectra $P_0$, $P_2$ and $P_4$ for H$\alpha$, [OIII] and [OII] at $z=1$ with interloper lines are shown. The uncertainties are also shown in shaded regions by considering the uncertainties of SFR-line luminosity relations and SFRD. For comparison, both projected and original power spectra of the interloper lines are shown in dash-dotted and dashed curves, respectively. For H$\alpha$ observation at $z=1$ (left top and bottom panels of Figure~{\ref{fig:Pl_err}}), we assume there is no strong interloper line can contaminate H$\alpha$ signal significantly \citep{Gong17}. We find that the shot noise term (red dotted line) will not affect the total observed $P_0$ (red solid curve) in the linear regime $k\lesssim0.1$ ${\rm Mpc}^{-1}\, h$, which is also true for the  [OIII] and [OII] cases. In the [OIII] observation at $z=1$ (middle top and bottom panels of Figure~{\ref{fig:Pl_err}}), the projected H$\alpha$ power spectra from $z=0.53$ (red dash-dotted) are stronger than [OIII] by a factor of $3-4$ at large scales, and can be considerably affect the [OIII] measurement. This is the same for the [OII] case (right top and bottom panels of Figure~{\ref{fig:Pl_err}}), the projected H$\alpha$ from $z=0.14$ (red dash-dotted) and [OIII] from $z=0.49$ (orange dash-dotted) can significantly contaminate the [OII] measurement at $z=1$. On the other hand, the H$\beta$ from $z=0.53$ (green dash-dotted) seems too weak to affect the result.

\begin{figure*}
\centerline{
\includegraphics[scale=0.27]{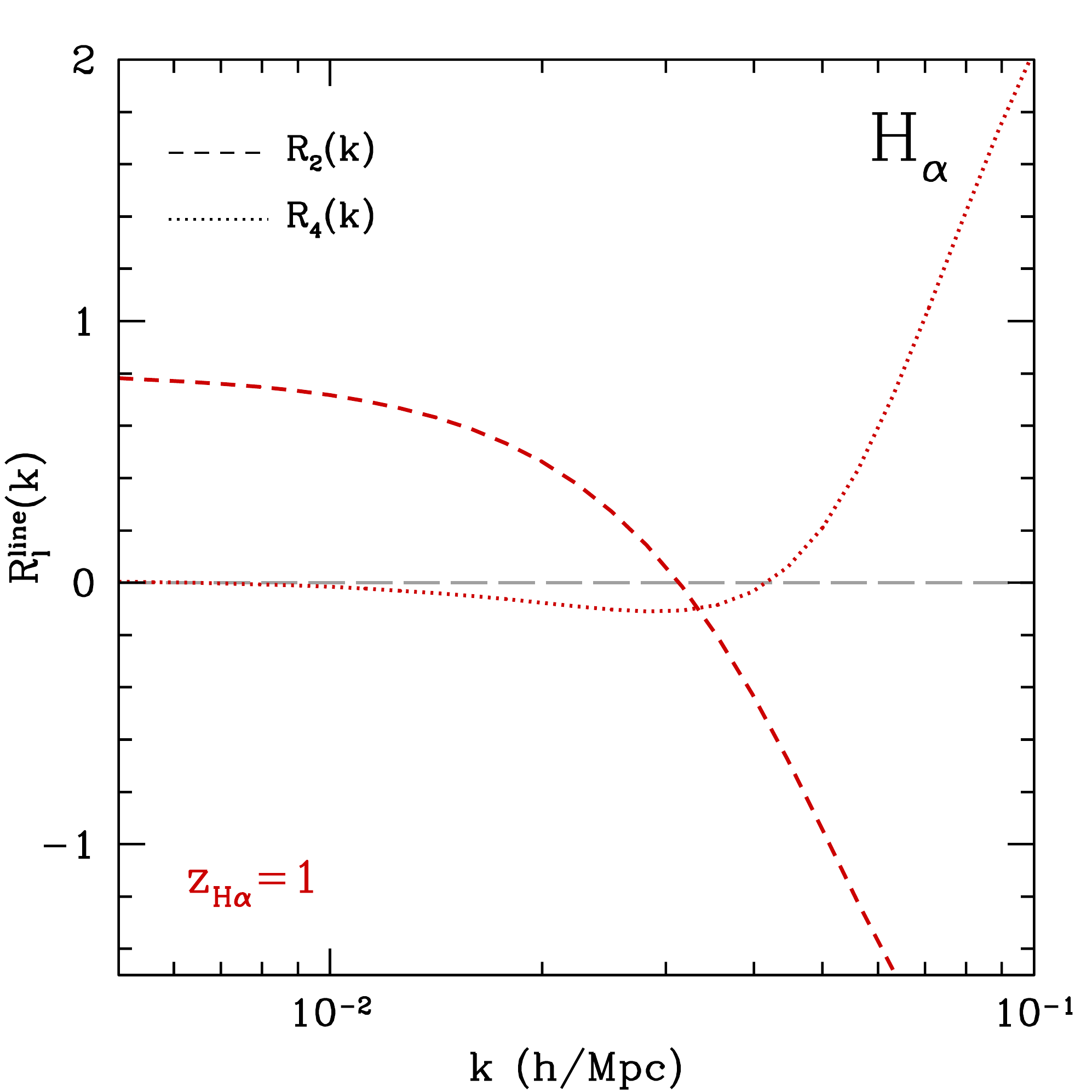}
\includegraphics[scale=0.27]{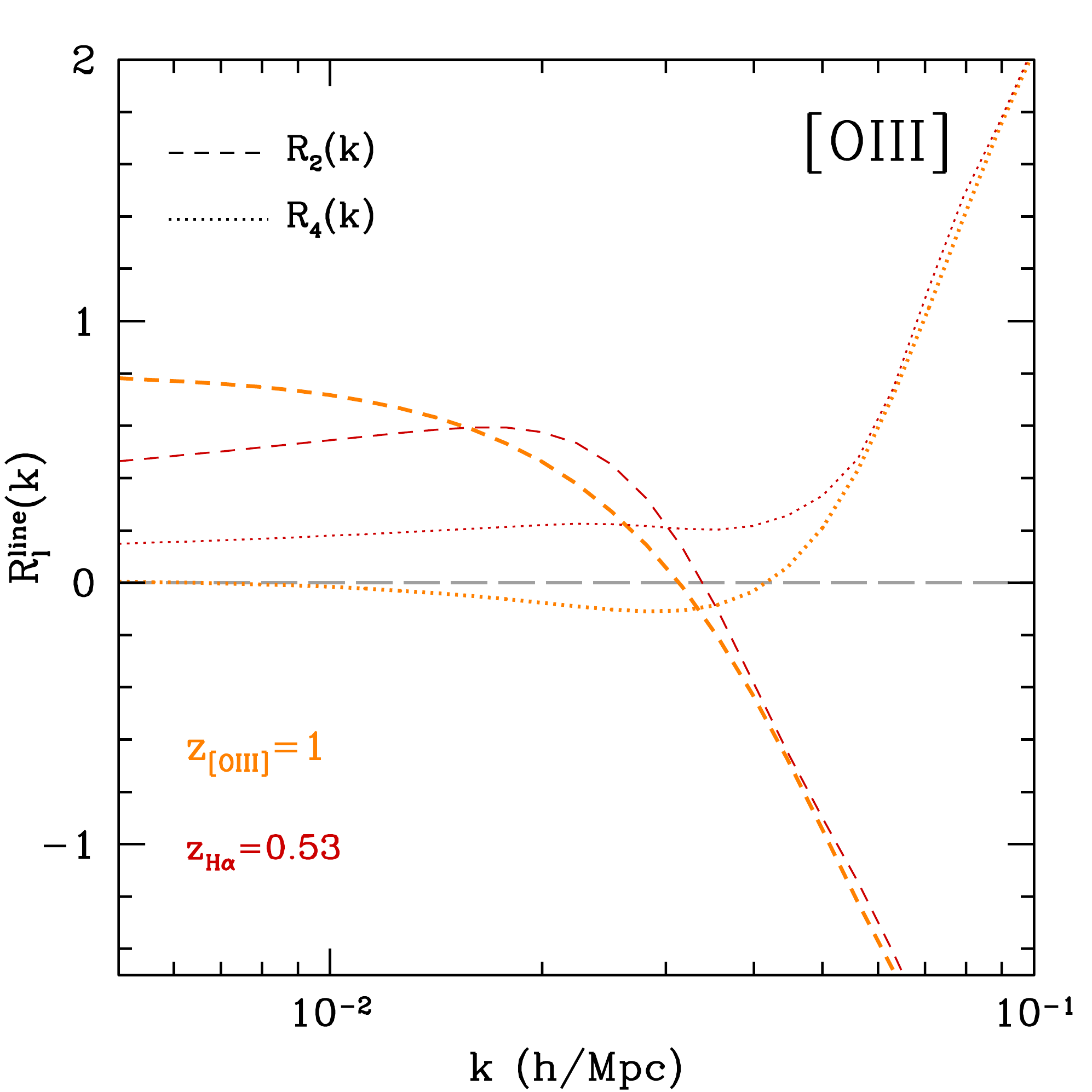}
\includegraphics[scale=0.27]{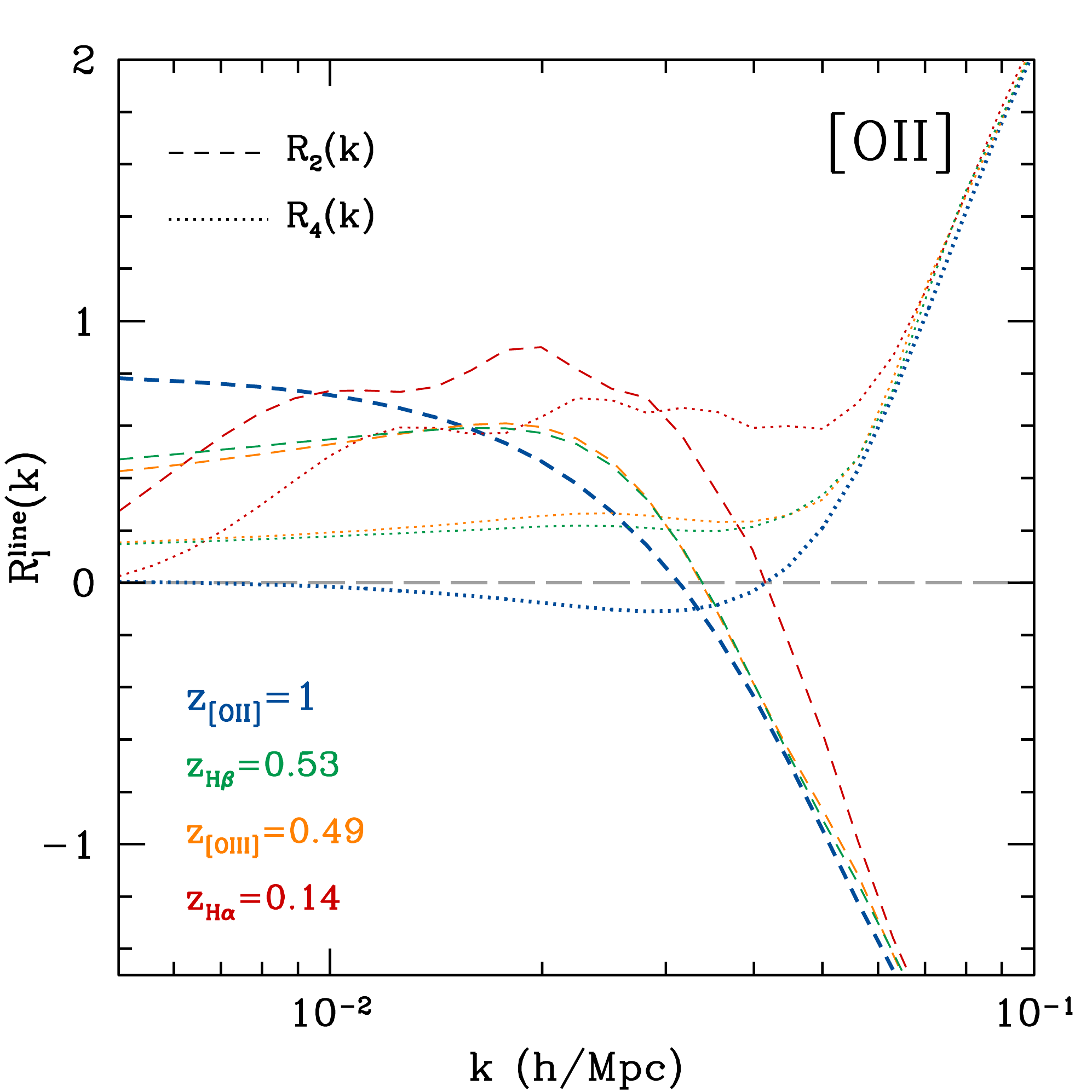}}
\caption{The ratios of $P_2$ and $P_4$ to $P_0$, i.e. $R_2$ and $R_4$ for the signal and interloper lines. The dashed and dotted curves are for $R_2$ and $R_4$, respectively. The H$\alpha$, [OIII], [OII] and H$\beta$ curves are in red, orange, blue and green, respectively. We can find that the shapes of $R_2$ and $R_4$ of the signal lines are different from the interloper lines, which can help to distinguish the interlopers from the signals.}
\label{fig:ratio_lines}
\end{figure*}

In order to remove or reduce the contaminations of the interloper lines, we try to distinguish them by the differences of features on the multipole power spectra between signal and interlopers. Following \cite{Lidz16}, we calculate the ratio of $P_2$ and $P_4$ to $P_0$ for H$\alpha$, [OIII] and [OII] lines at $z=1$ and their interlopers as shown in Figure~{\ref{fig:ratio_lines}}. We define $R_2(k)=P_2(k)/P_0(k)$ and $R_4(k)=P_4(k)/P_0(k)$. We can find that the shapes of $R_2(k)$ and $R_4(k)$ of the interloper lines are different from that of the signal lines. For $R_2$ curves, unlike continuously declines for the signal line, the interloper curves first rise up and then decrease around $k=0.02$ ${\rm Mpc}^{-1}\, h$. In the $R_4$ case, they are always positive for the interloper lines we study, while it is less than 0 at large scales at $k\lesssim0.05$ ${\rm Mpc}^{-1}\, h$ for the signal line. Besides, as can be seen, especially for the [OII] case (right panel of Figure~{\ref{fig:ratio_lines}}), the wider of the redshift intervals between the signal and interlopers, the larger of differences of their $R_2(k)$ and $R_4(k)$. This is quite useful for distinguishing the interlopers, since the projection effect become stronger for larger redshift intervals between the signal and interloper lines (e.g. see the $P_0^{\rm [OII]}$ case by comparing the dash-dotted and dashed curves shown in the right top panel of Figure~{\ref{fig:Pl_err}}). This implies that although the contamination effect could be larger for interloper lines from lower redshift, but they should be easier to be identified by comparing their $R_2$ and $R_4$ to the signal line, or  using all information from $P_0$, $P_2$ and $P_4$. This is also the concept we adopt to deal with the interloper lines in this work.

We should also note that there could be more information mined in the full RSD power spectrum $P(k,\mu)$ beyond the multipole moments $P_0$, $P_2$ and $P_4$, considering the AP effect and interloper lines. This means that the $P(k,\mu)$ data could potentially provide more stringent constraints on the cosmological and astrophysical parameters, although it may be not as efficient as the multipole moments. We will explore this issue quantitively in our future work.

\section{Line Detection}
\label{sec:LD}

\begin{figure*}
\centerline{
\includegraphics[scale=0.27]{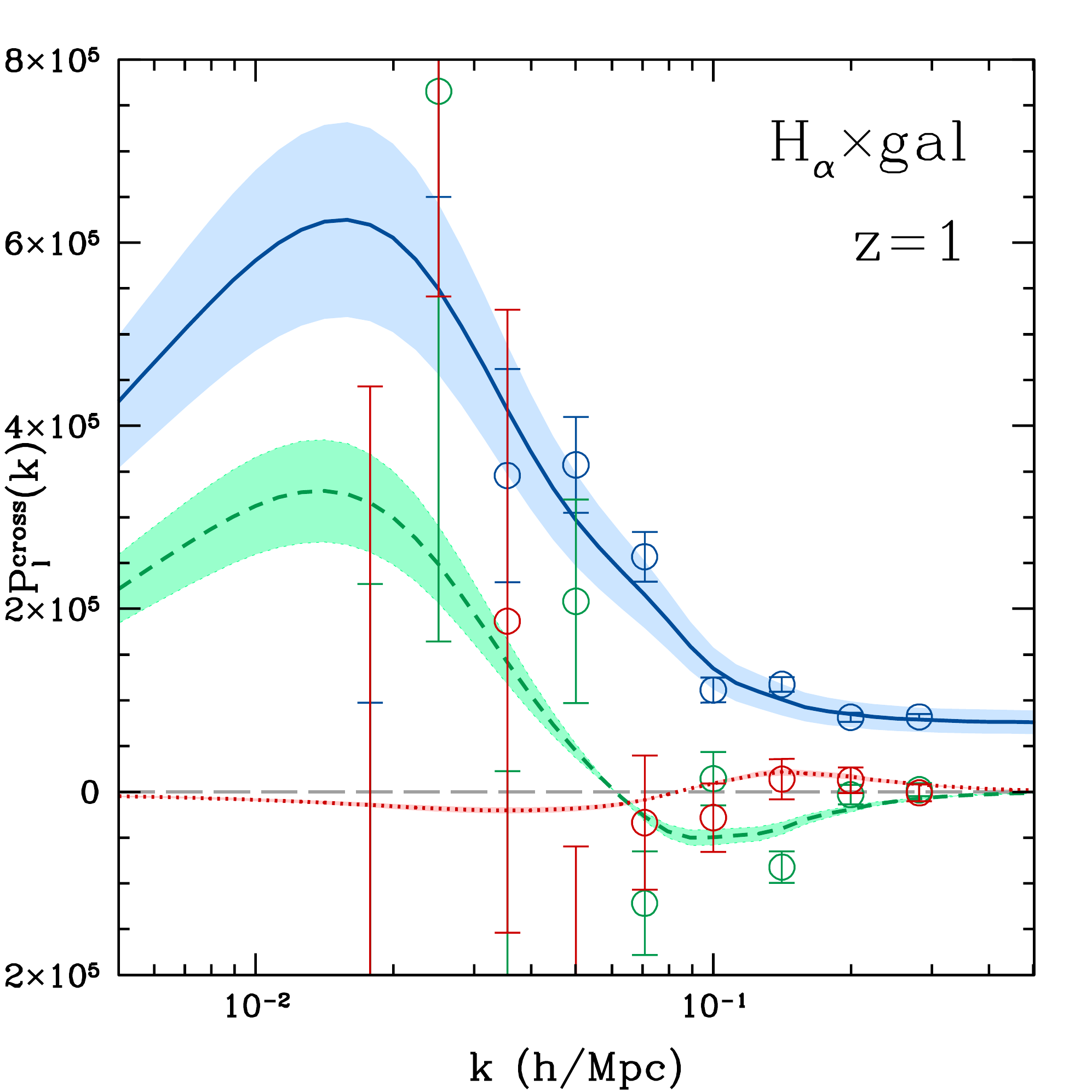}
\includegraphics[scale=0.27]{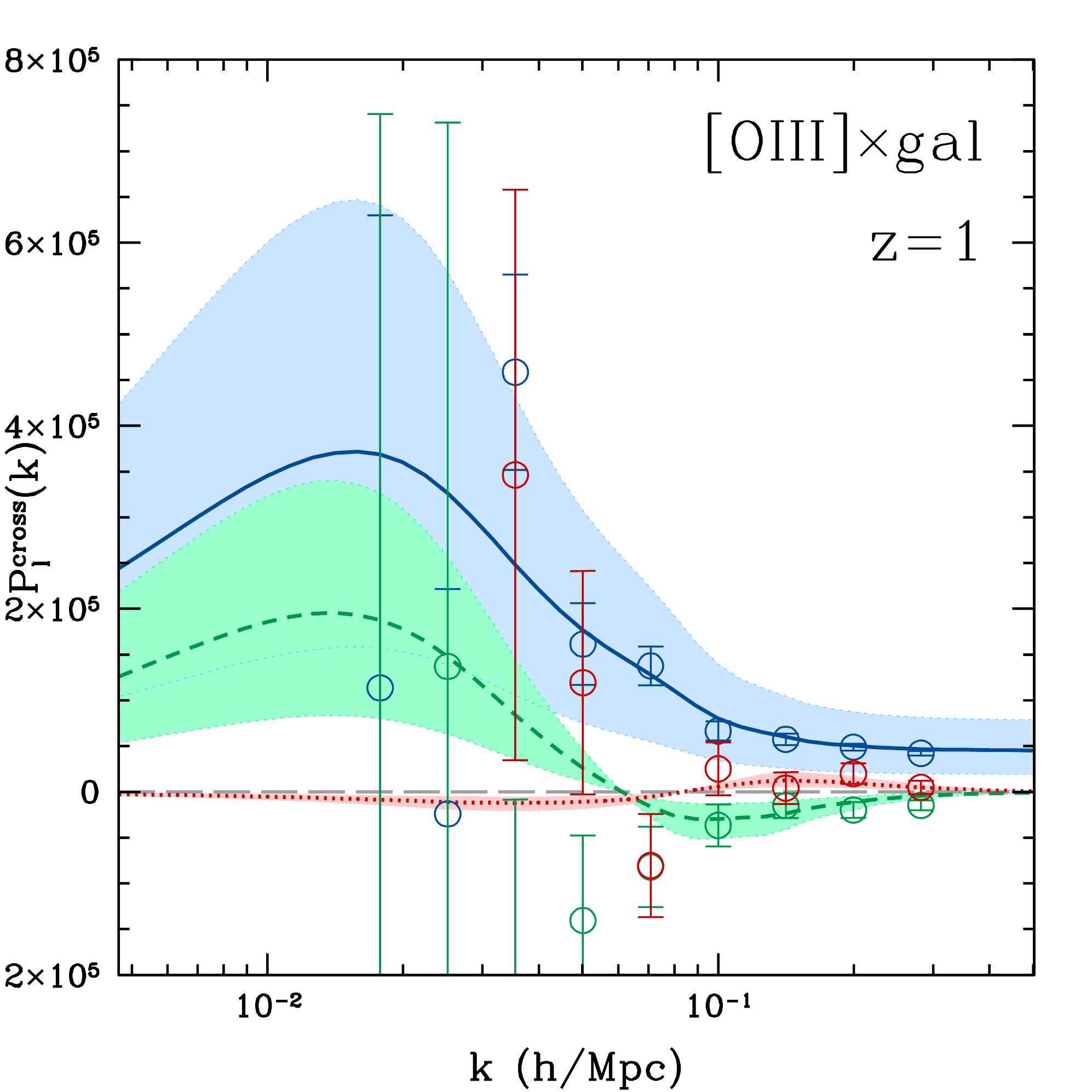}
\includegraphics[scale=0.27]{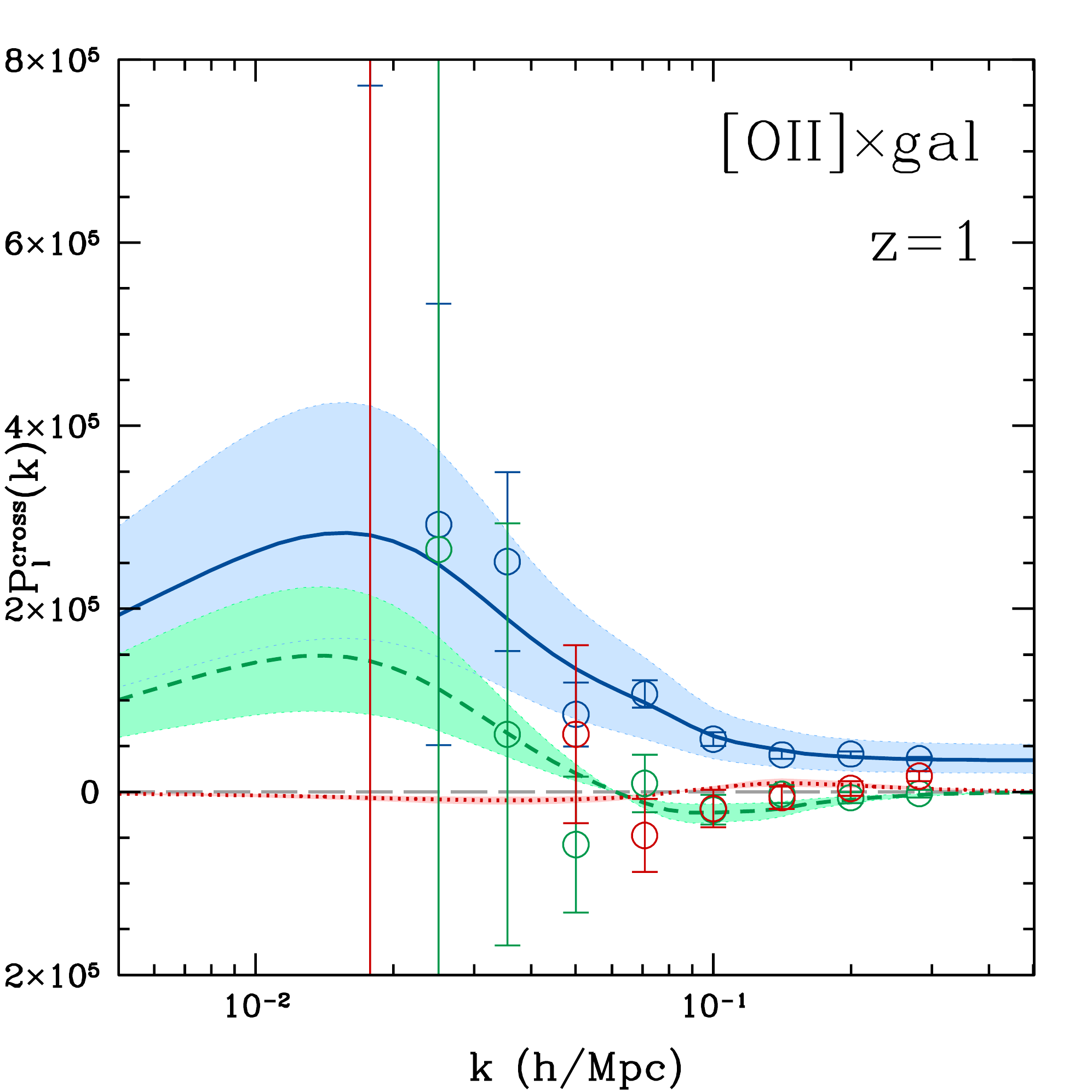}}
\caption{The multipole moments of cross power spectra of H$\alpha$, [OIII] and [OII] and CSST spectroscopic galaxy survey at $z=1$ are shown in the left, middle and right panels, respectively (in units of $\rm (Jy/sr)\,(Mpc/h)^3$). The blue solid, green dashed, and red dotted curves denote the $P_0$, $P_2$ and $P_4$ terms, respectively. The shaded regions show the uncertainties of the power spectra derived from errors of SFRD and line luminosity-SFR relations as discussed in Section~\ref{sec:LIM}.}
\label{fig:Pl_cross}
\end{figure*}

Here we adopt SPHEREx experiment to perform the measurements\footnote{\tt http://spherex.caltech.edu/}. SPHEREx is a proposed near-infrared space telescope that exploring from 0.75 to 5 $\mu$m \citep{Dore14, Dore16, Dore18}. It has a diameter of 20 cm, and can obtain spectra with 6.2$\times$6.2 arcsec$^2$ pixel size. The spectral resolutions are different in its four bands, that we have $R=41$ in $0.75<\lambda<2.42\ \rm \mu m$, $R=35$ in $2.42<\lambda<3.82\ \rm \mu m$, $R=110$ in $3.82<\lambda<4.42\ \rm \mu m$, and $R=130$ in $4.42<\lambda<5.00\ \rm \mu m$. In our study, we only consider the first two bands for measuring H$\alpha$, [OIII] and [OII] lines at $z\lesssim3$. We explore the detectability of the lines using its deep survey within 200 deg$^2$. 

The covariance matrix of observed line power spectrum at a given redshift can be estimated by \citep[see e.g.][]{Chung19}
\ba \label{eq:Del_Pk}
&{\rm Cov}&[P_{\ell,{\rm obs}}(k,z),P_{\ell',{\rm obs}}(k,z)] = \frac{(2\ell+1)(2\ell'+1)}{N_{\rm m}(k,z)} \nonumber\\
               &\times& \int_0^1 d\mu\, \mathcal{L}_{\ell}(\mu) \mathcal{L}_{\ell'}(\mu) [P_{\rm obs}(k,\mu,z)+P_{\rm N}(z)]^2,
\ea
where $P_{\rm obs}(k,\mu,z)$ is the observed redshift-space total line power spectrum, which is composed of signal and projected interloper power spectra. The error of $P_{\ell,{\rm obs}}(k,z)$ is then can be calculated by $\Delta P_{\ell,{\rm obs}}(k,z)=\sqrt{{\rm Cov[P_{\ell,{\rm obs}}(k,z),P_{\ell,{\rm obs}}(k,z)]}}$ where $\ell=\ell'$. $P_{\rm N}(z)$ is the noise power spectrum determined by instrumental noise. It is given by
\be
P_{\rm N}(z)=V_{\rm pix}(z)\frac{\sigma_{\rm pix}^2}{t_{\rm pix}}.
\ee
Here $V_{\rm pix}(z)$ is the pixel volume at $z$ which can be calculated by the SPHEREx spatial and frequency resolutions, and $\sigma_{\rm pix}^2/t_{\rm pix}$ is the squared instrument thermal noise per survey pixel, where $t_{\rm pix}$ denotes the integration time per pixel.  We adopt 2.2, 3.1 and 3.9 $\rm nW\,m^{-2}sr^{-1}$ for SPHEREx Ha, [OIII] and [OII] surveys at z=1, respectively. $N_{\rm m}(k,z)$ is the number of Fourier modes in a wavenumber interval $\Delta k$ at $k$ in the upper-half wavenumber plane. As an approximation, it can be estimated by
\be \label{eq:Nm}
N_{\rm m}(k,z) = 2\pi k^2\Delta k\,\frac{V_{\rm S}(z)}{(2\pi)^3},
\ee
where $V_{\rm S}$ is the total survey volume at $z$. In practice, we perform a real counting of the modes to obtain an exact $N_{\rm m}$ in each wavenumber interval. This can avoid discrepancy between the real $N_{\rm m}$ and the one given by Eq.~(\ref{eq:Nm}), especially at small scales (large $k$). Then the signal to noise ratio (SNR) can be calculated by
\be \label{eq:SNR}
{\rm SNR}(z) = \sqrt{\sum_{k\,\rm bin}\left[\frac{P_{\ell,\rm obs}(k,z)}{\Delta P_{\ell,\rm obs}(k,z)}\right]^2}.
\ee
Here the $k$ range is determined by survey volume and spatial and frequency resolution. Considering the spatial and frequency resolutions of SPHEREx and avoiding non-linear effect, we take $0.01<k<0.3\ {\rm Mpc^{-1}}h$ in our analysis. Note that the low $k_{\parallel}$ modes can be lost due to instrumental and foreground contaminations in intensity mapping surveys, which can impact constraints on the parameters that are sensitive to low $k$ modes, such as $f_{\rm NL}$ \citep{Dizgah19}.

In Figure~\ref{fig:Pl_err}, we show the error $\Delta P_{\ell,{\rm obs}}$ of the observed total multipole power spectra $P_0$ (upper panels), $P_2$, and $P_4$ (bottom panels) for H$\alpha$, [OIII] and [OII] lines at $z=1\pm0.2$. According to the covariance matrix derived from Eq.~(\ref{eq:Del_Pk}), a random shift from Gaussian distribution is added in each data point. As can be seen, we can obtain good measurements on total power spectra of each line, and we have SNR=10.4, 7.7 and 5.8 for H$\alpha$ line, 12.7, 9.1 and 6.9 for [OIII] line, and 19.1, 9.5 and 7.2 for [OII] line, for $P_0$, $P_2$, and $P_4$, respectively. Although the signal power spectra of [OII] (blue dashed curve in the top-right panel) and [OIII] (orange dashed curve in the top-middle panel) are lower than H$\alpha$ (red shot-long dashed curve in the top-left panel), we find that they have larger SNR, since they suffer strong contaminations (dash-dotted curves) from other lines at lower redshifts which can boost up their total power spectra. 

Note that higher SNR with higher amplitude of total power spectrum does not mean we can obtain more stringent constraints on the cosmological and astrophysical parameters we are interested in. Since we will consider both signal and interlopers in our fitting process, high total power spectrum, such as [OII], has more components from interlopers with more free parameters. As we discuss in Section~\ref{sec:results}, it is possible that this can somehow loose the constraint on some paramters.

\section{cross-correlation with galaxy survey}
\label{sec:CC}

\begin{figure*}
\centerline{
\includegraphics[scale=0.27]{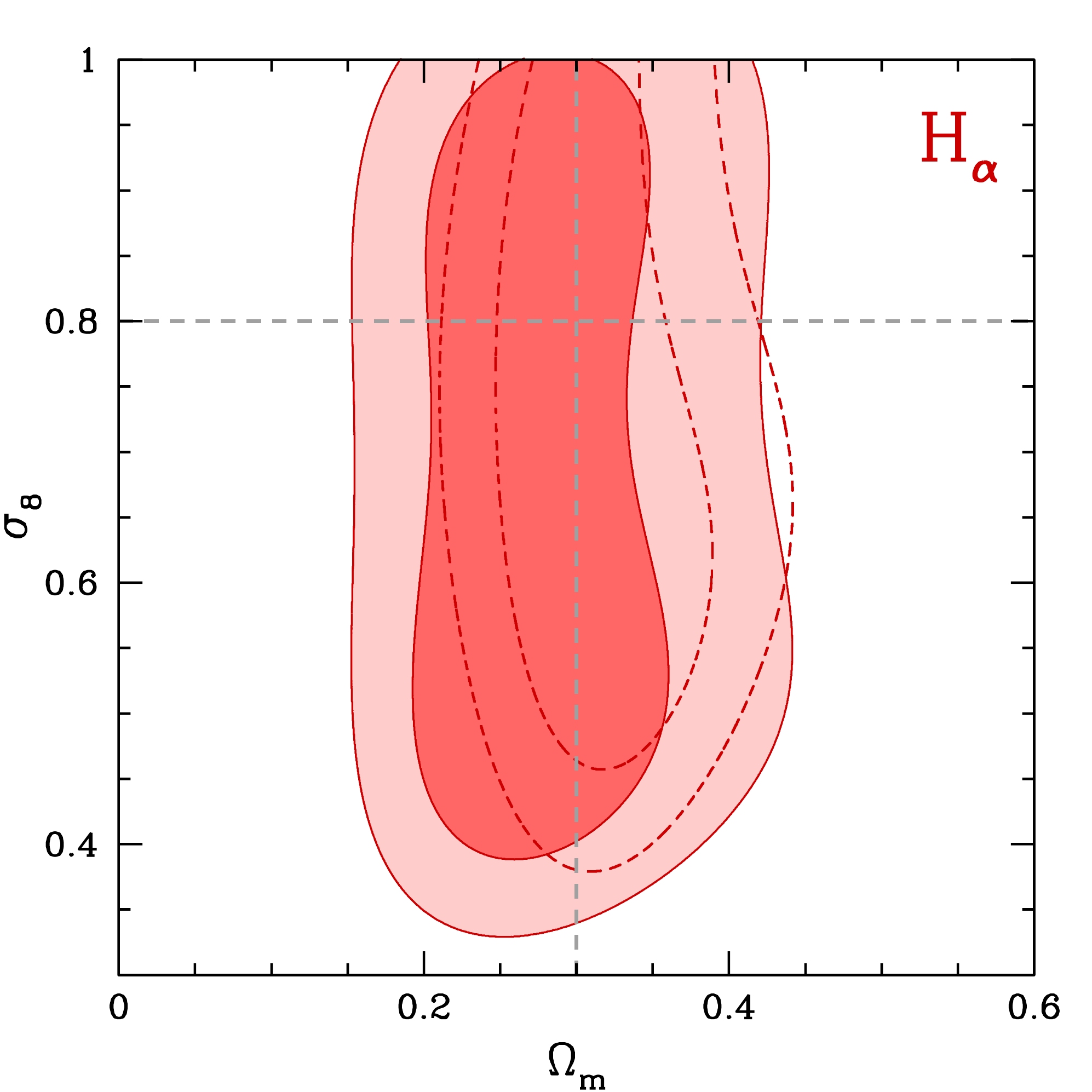}
\includegraphics[scale=0.27]{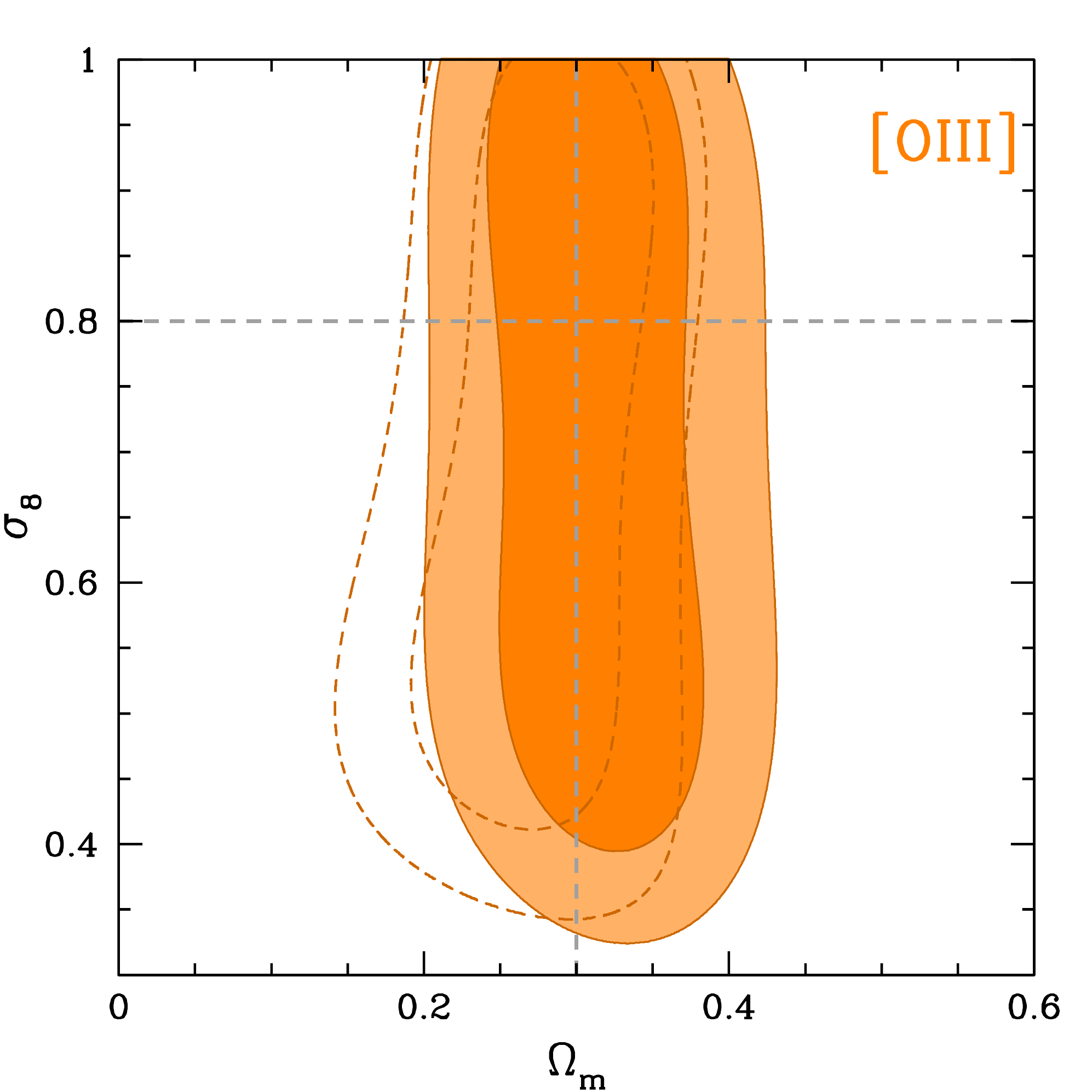}
\includegraphics[scale=0.27]{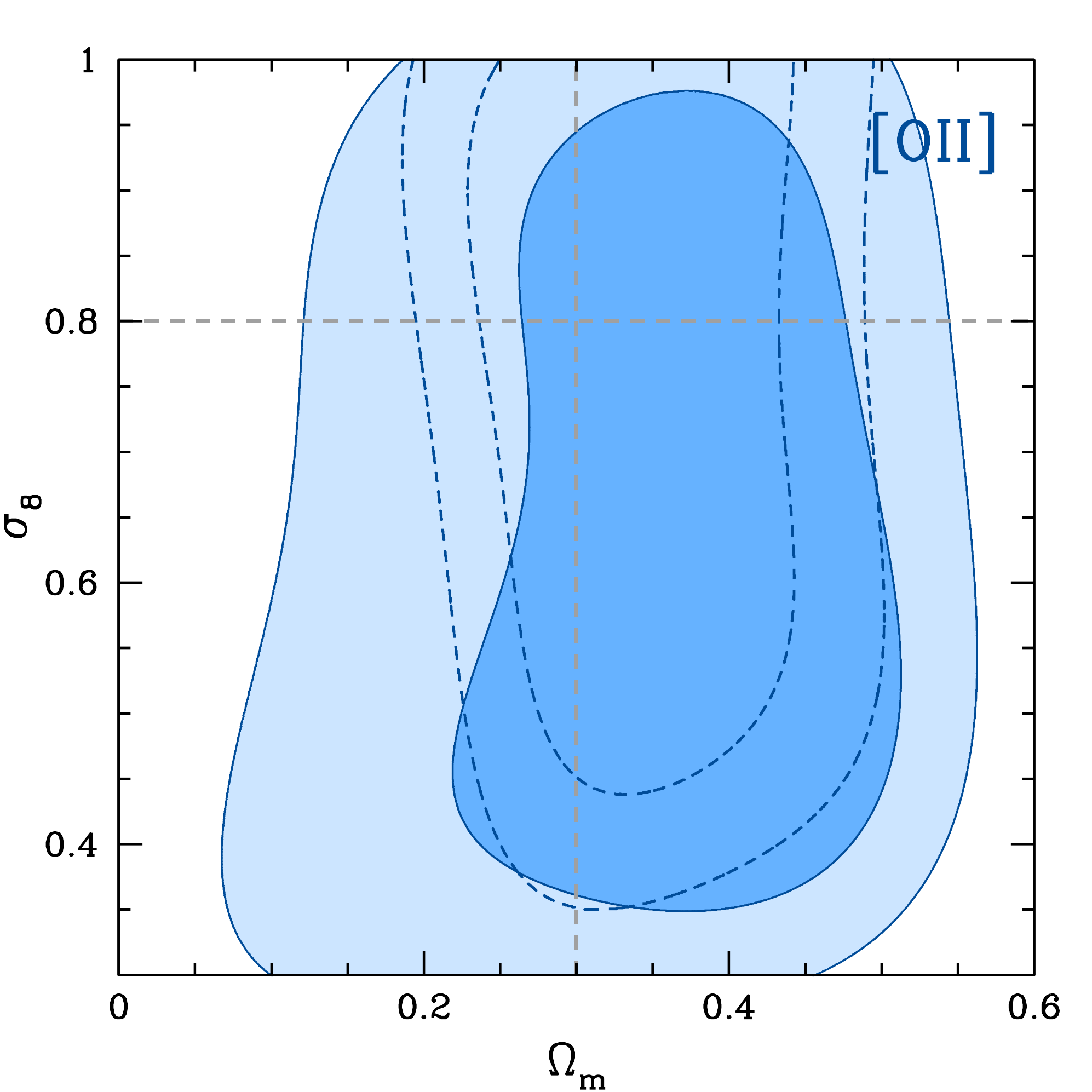}}
\centerline{
\includegraphics[scale=0.27]{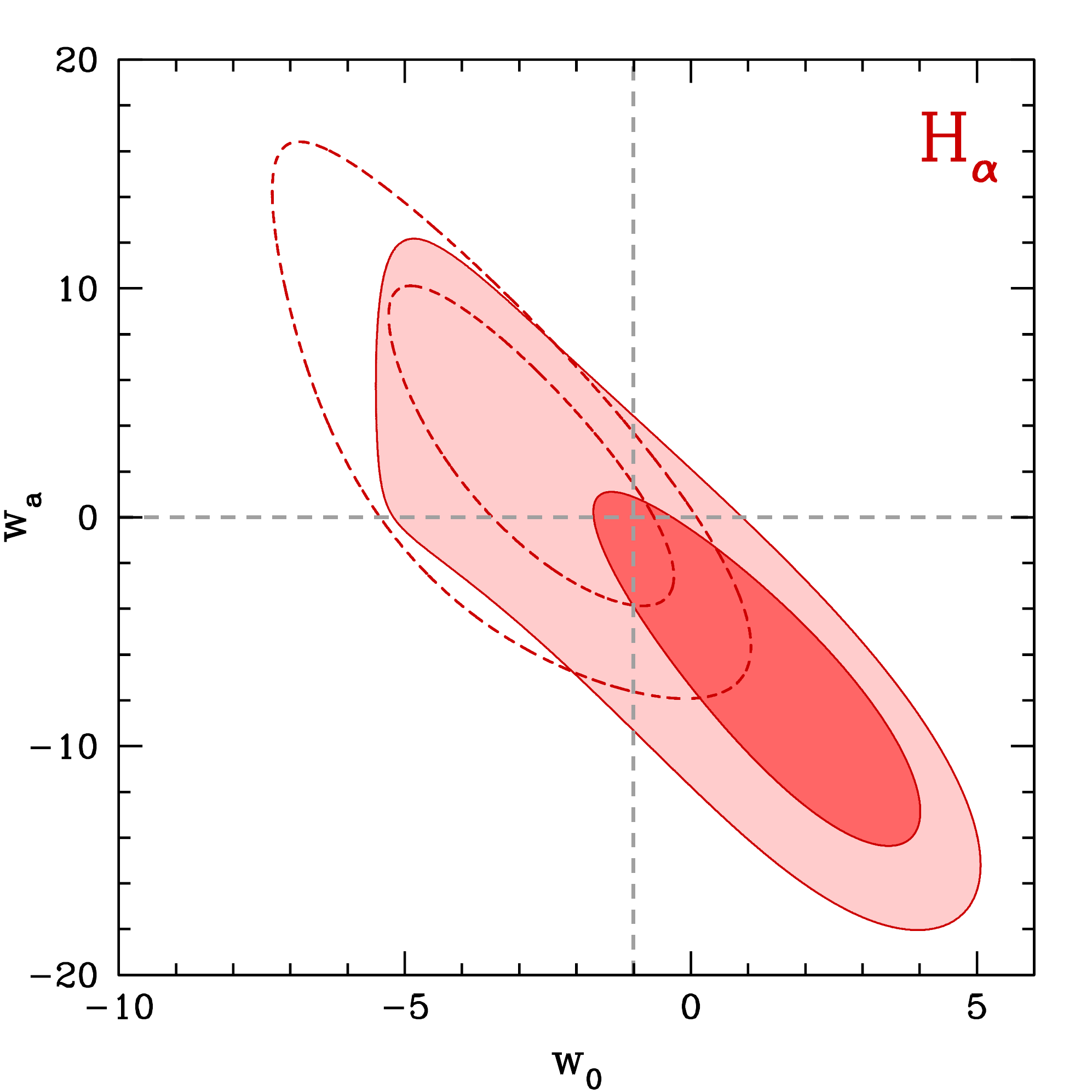}
\includegraphics[scale=0.27]{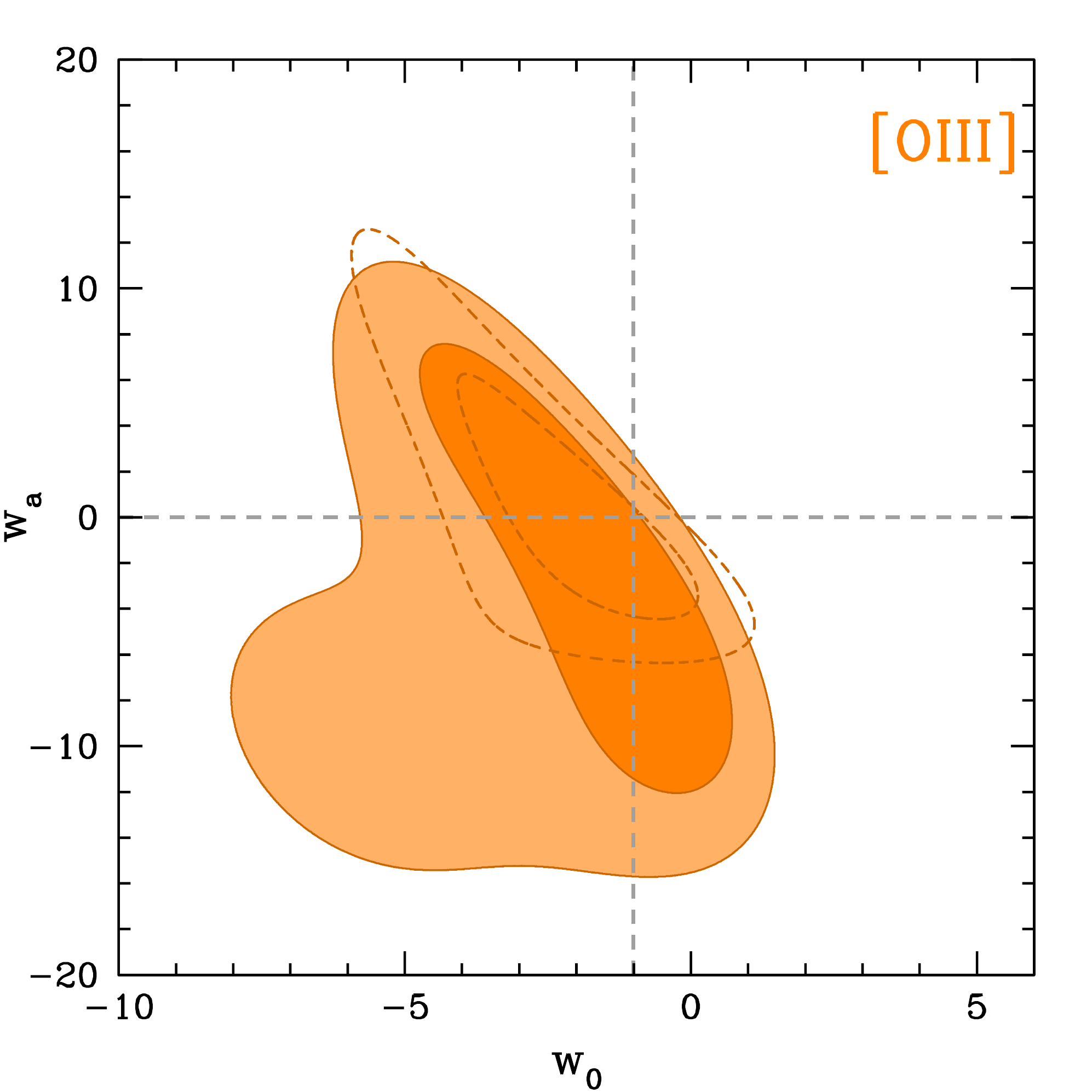}
\includegraphics[scale=0.27]{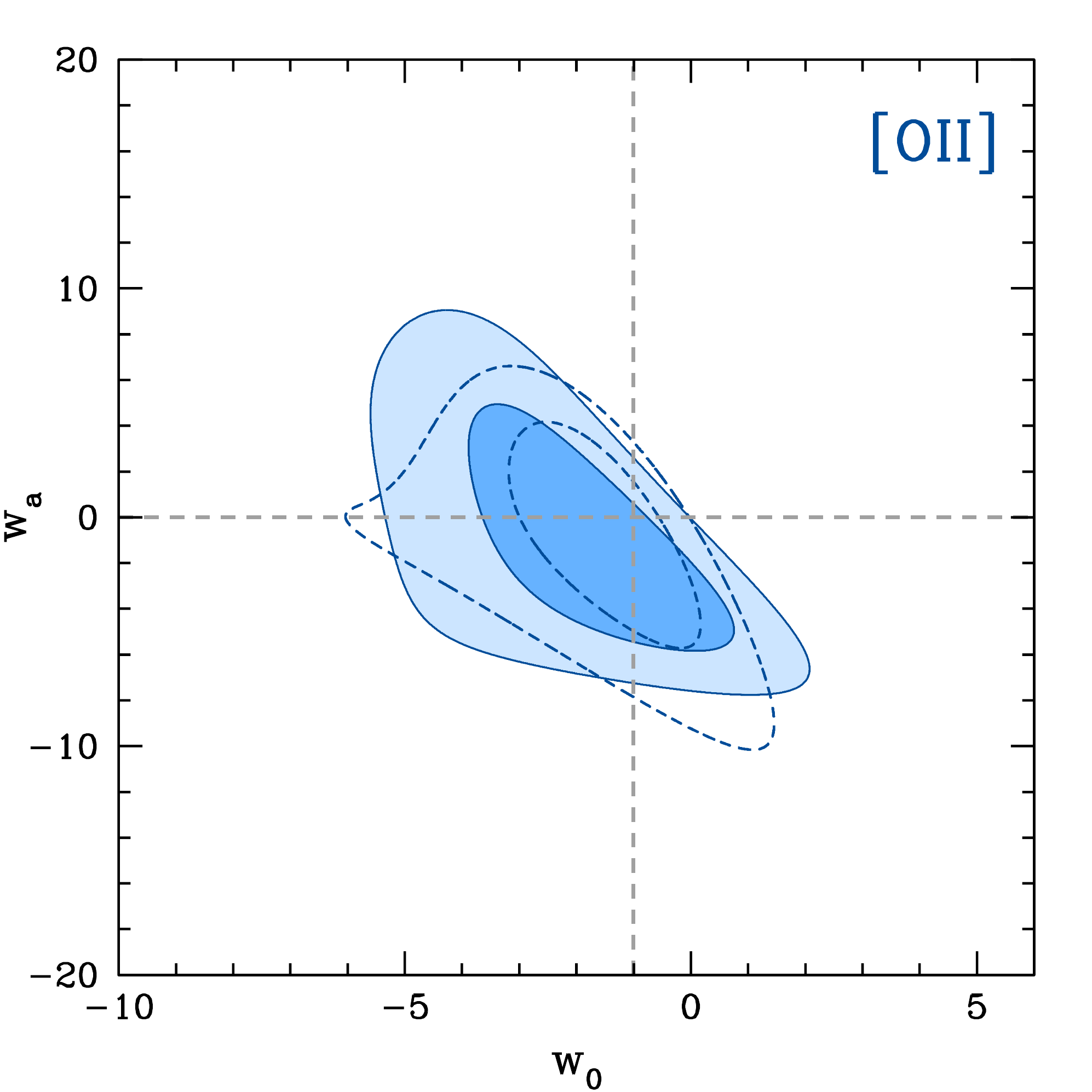}}
\caption{The contour maps of $\Omega_{\rm M}$ vs. $\sigma_8$ (upper panels) and $w_0$ vs. $w_a$ (lower panels) for 1$\sigma$ (68.3\%) and 2$\sigma$ (95.5\%) C.L. from the constraints of H$\alpha$ (left), [OIII] (middle) and [OII] (right) intensity mapping. The solid and dashed contours denote the line auto power spectra only that measured by SPHEREx experiment and cross power spectrum included with CSST spectroscopic galaxy survey, respectively. The gray dashed lines indicate the fiducial values of the parameters.}
\label{fig:m0s8_w0wa_lines}
\end{figure*}

\begin{figure*}
\centerline{
\includegraphics[scale=0.27]{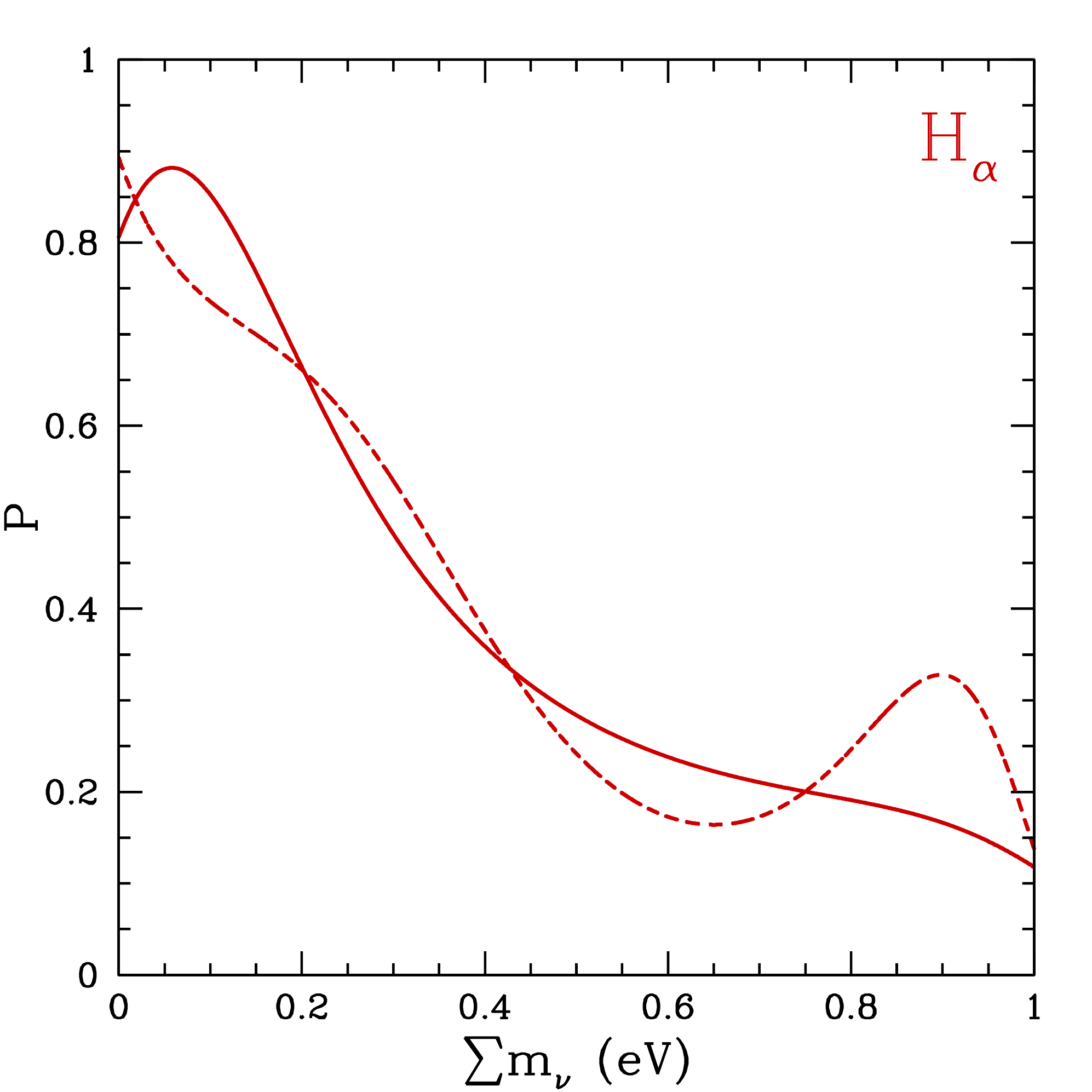}
\includegraphics[scale=0.27]{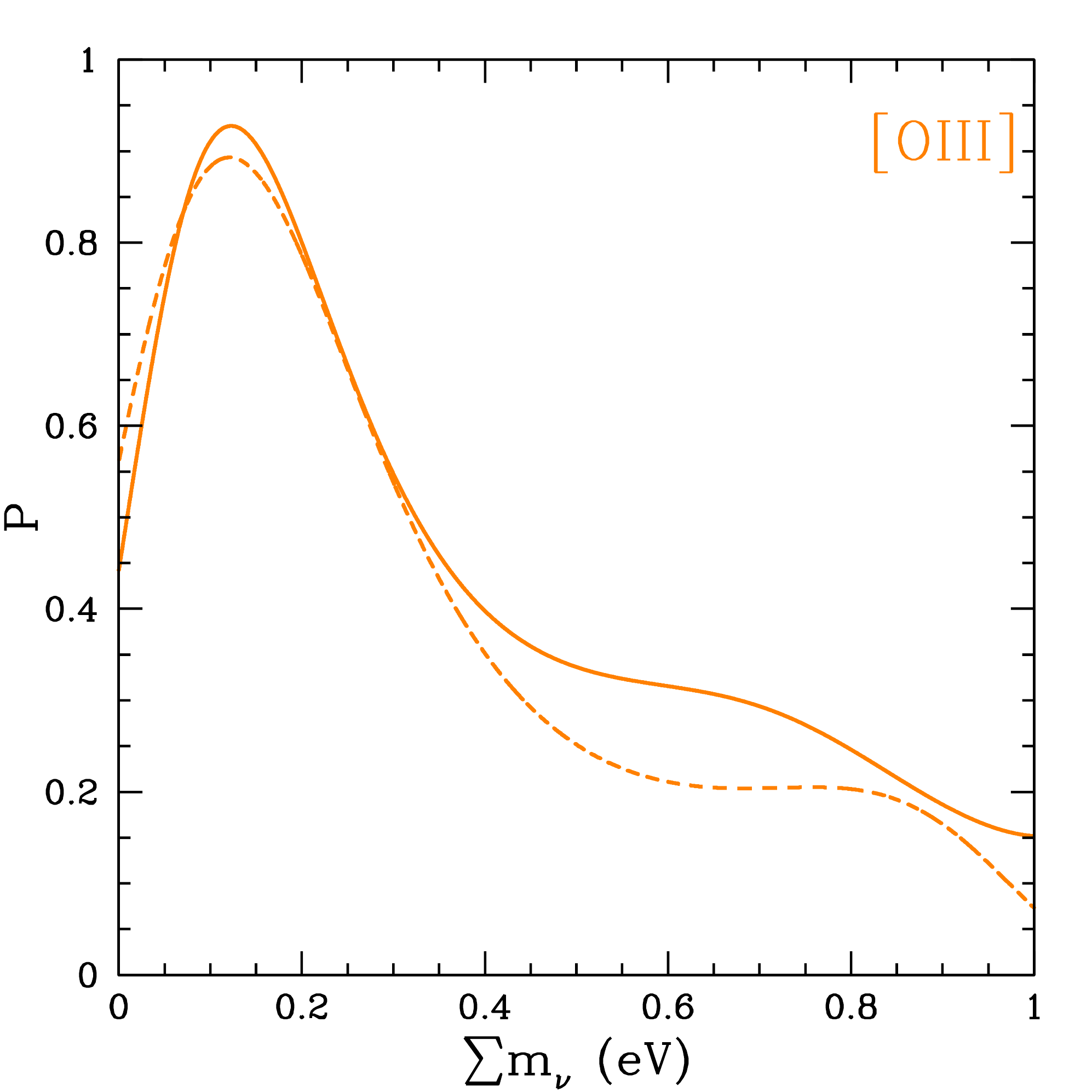}
\includegraphics[scale=0.27]{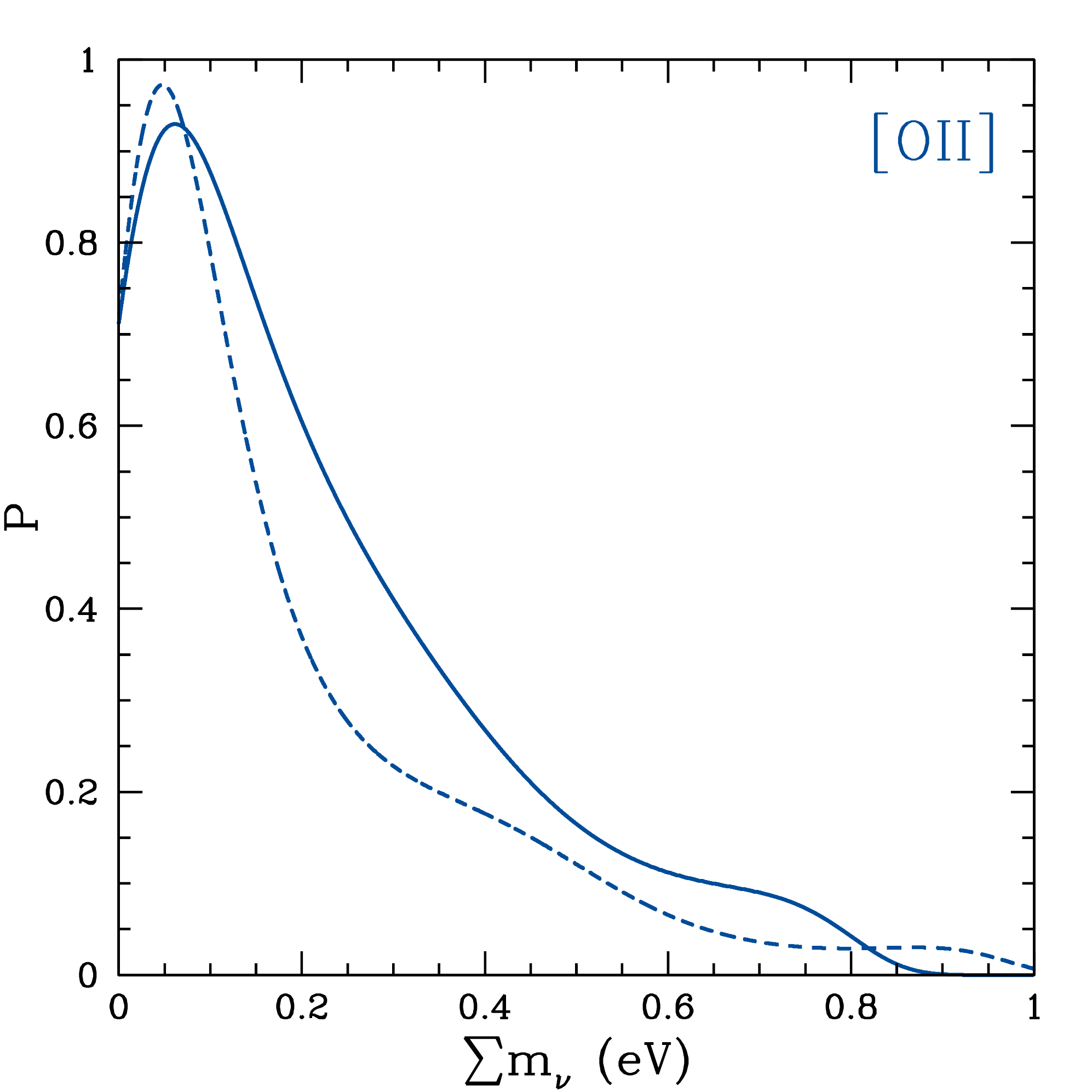}}
\centerline{
\includegraphics[scale=0.27]{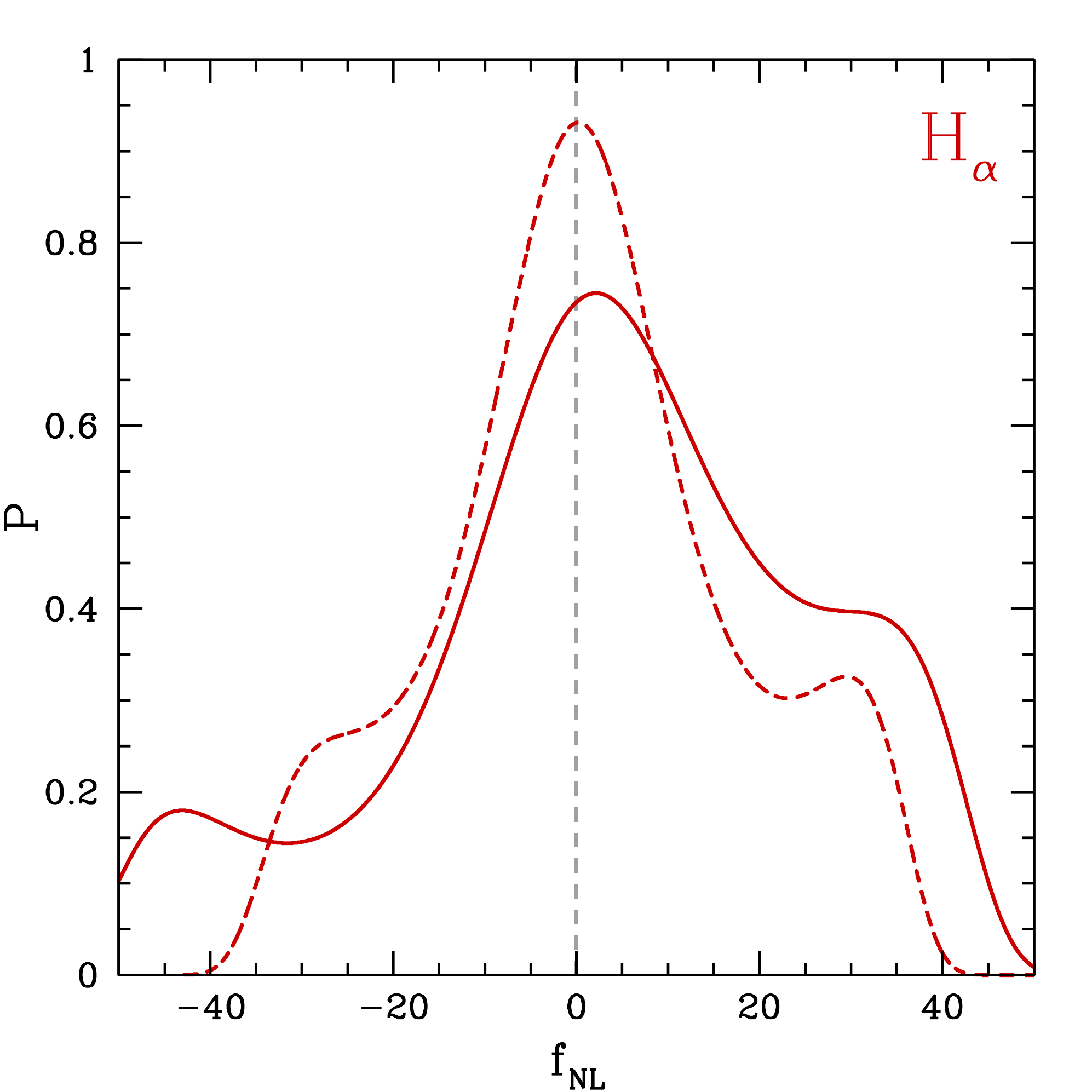}
\includegraphics[scale=0.27]{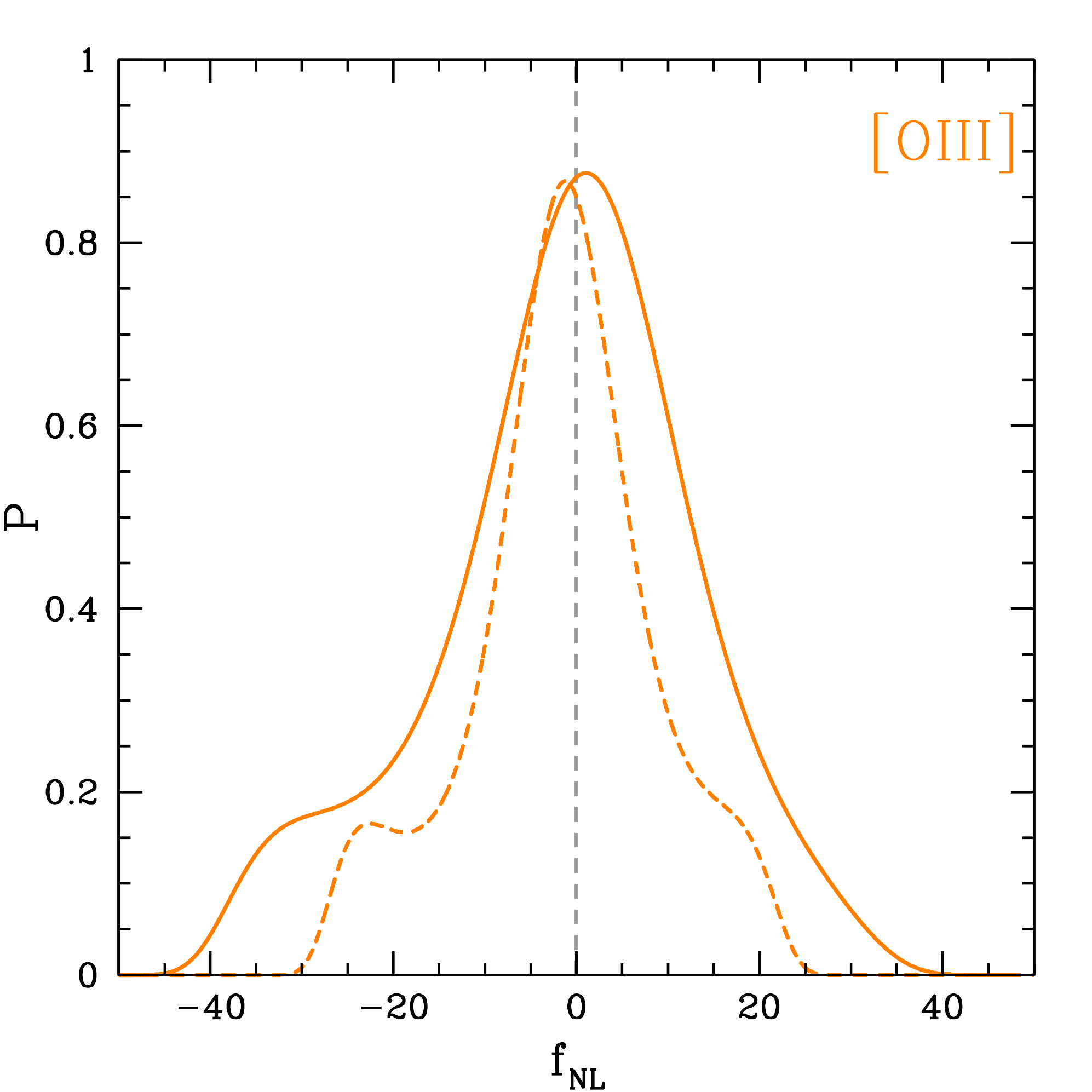}
\includegraphics[scale=0.27]{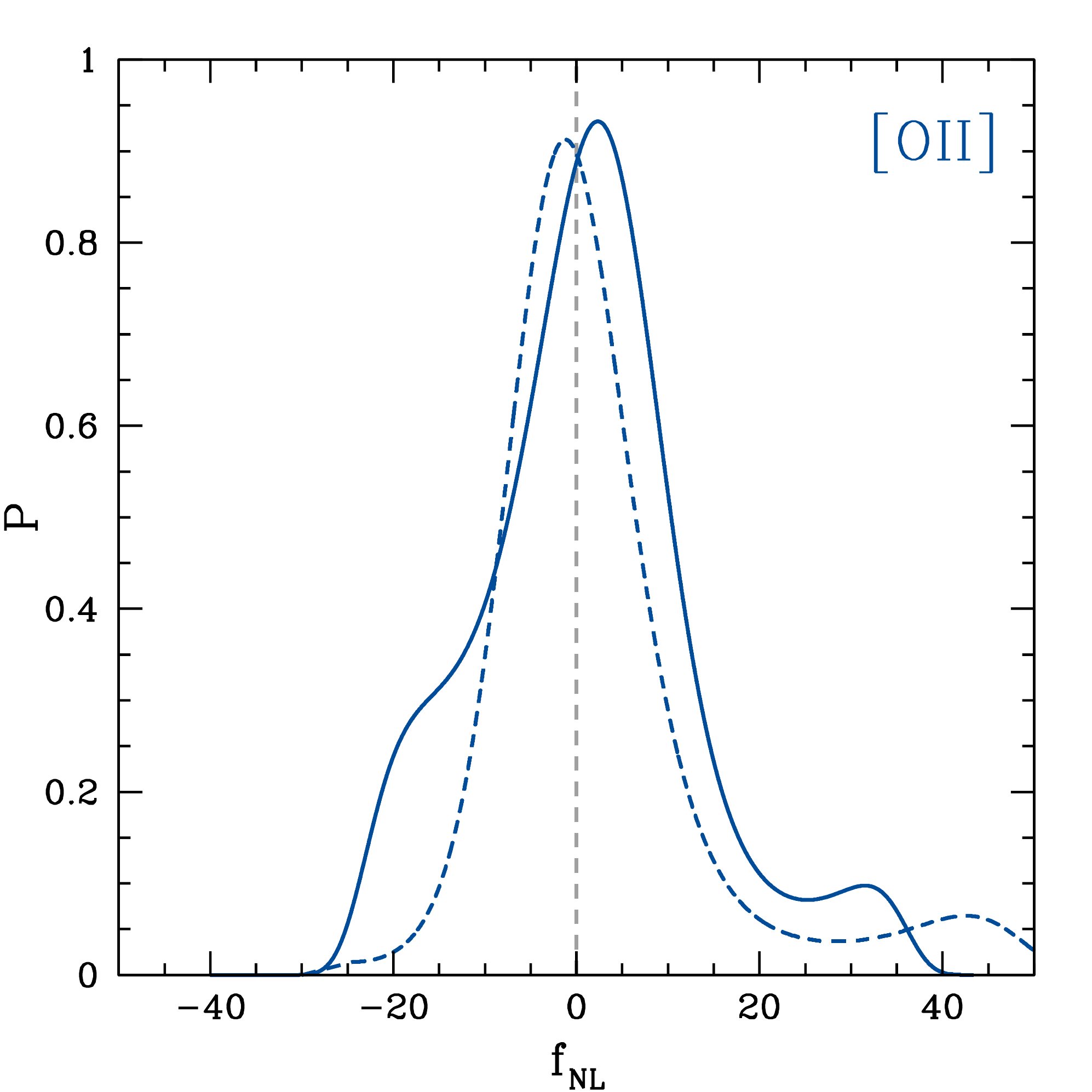}}
\caption{The 1D PDFs of $\sum m_{\nu}$ and $f_{\rm NL}$ derived from the constraints by H$\alpha$ (left), [OIII] (middle) and [OII] (right) intensity mapping. The solid and dashed curves denote the results from fitting line auto power spectra only and cross power spectra included cases, respectively.}
\label{fig:mv_fnl_lines}
\end{figure*}

An effective way to reduce contaminations is cross-correlating intensity mapping survey with other kinds of surveys, such as galaxy survey \citep[see e.g.][]{Chang10}. In this work, we take CSST spectroscopic galaxy survey for discussion. CSST is a two meter space telescope established by the space application system of the China Manned Space Program \citep{Zhan11, Zhan18, Cao18, Gong19}. It has seven photometric imaging and three slitless spectroscopic bands ranging from 255 to 1000 nm, and will simultaneously cover about $17,500$ deg$^2$ with a field of view 1.1 deg$^2$. The magnitude limit can reach $r\sim26$ AB mag for 5$\sigma$ point source detection in the photometric survey, and $\sim23$ for the spectroscopic survey. The galaxy distribution has a peak around $z=0.7$ and 0.3, and can extend to as high as $z=5$ and 2 for its photometric and spectroscopic surveys, respectively. In particular, the CSST spectroscopic survey has a spectral resolution $R\gtrsim200$, and can obtain a galaxy number density $n_{\rm gal}\sim 5\times10^{-3}$ $({\rm Mpc}/h)^{-3}$ at $z\sim 1$ \citep{Gong19}. 

The apparent redshift-space cross power spectrum of line intensity map and galaxy survey can be estimated by
\ba \label{eq:P_cross}
P^{\rm (s)}_{\rm cross}(k',\mu',z) &=& P_{\rm cross}^{\rm clus}(k',z) (1+\beta\mu'^2)(1+\beta_{g}\mu'^2) \nonumber \\
                                                 &\times& \mathcal{D_{\rm c}}(k',\mu',z) + P_{\rm cross}^{\rm shot}(z).
\ea
Here $\beta_{g}=f(z)/b_{g}(z)$, where $b_{g}$ is the galaxy bias \citep{Gong19}, and $\mathcal{D_{\rm c}}$ is the damping term at small scales for both line intensity and galaxy surveys. At the linear regime we are interested in, we find that this term actually cannot significantly affect the results. The apparent real-space clustering cross power spectrum is given by
\be
P_{\rm cross}^{\rm clus}(k',z) = \bar{b}_{\rm line}(z) b_{g}(z) \bar{I}_{\rm line}(z) P_{\rm m}(k',z).
\ee
Assuming that all galaxies observed in traditional survey have emission lines that can be detected in intensity mapping survey, the shot-noise term takes the form as
\be\label{eq:Pshot_cross}
P^{\rm shot}_{\rm cross}(z) = \frac{1}{n_{\rm gal}(z)}\int_{M_{\rm min}}^{M_{\rm max}} {\rm d}M \frac{{\rm d}n}{{\rm d}M} \left[\frac{L_{\rm line}}{4\pi D_{\rm L}^2}y(z)D_{\rm A}^2\right].
\ee
As we mentioned in \S \ref{subsec:sps}, the line mean bias needs to be replaced by a scale-dependent bias $\bar{b}_{\rm line}(k',z)$ when considering primordial non-Gaussianity, and a factor $(1+\Delta P_{\rm m}/P_{\rm m})$ should be multiplied on the matter power spectrum $P_{\rm m}$ for massive neutrinos included in the model. 

Then the multipole moments of cross power spectrum $P_{\ell}^{\rm cross}(k,z)$ can be obtained by replacing $P^{\rm (s)}_{\rm line}(k',\mu')$ by $P^{\rm (s)}_{\rm cross}(k',\mu')$ in Eq.~({\ref{eq:Pl_line}}). The covariance of the multipole moments of cross power spectrum can be estimated by
\ba \label{eq:Del_Pcross}
&&{\rm Cov}[P_{\ell}^{\rm cross}(k,z),P_{\ell'}^{\rm cross}(k,z)] = \frac{(2\ell+1)(2\ell'+1)}{2\,N_{\rm m}^{\rm cross}(k,z)} \nonumber\\
               &\times& \int_0^1 d\mu\, \mathcal{L}_{\ell}(\mu) \mathcal{L}_{\ell'}(\mu) [P_{\rm cross}^2(k,\mu,z) + P_{\rm tot}^{\rm line}\,P_{\rm tot}^{\rm gal}],
\ea
Here $P_{\rm tot}^{\rm line}(k,u,z)=P_{\rm obs}^{\rm line}(k,u,z)+P_{\rm N}(z)$ is the total redshift-space line power spectrum, and $P_{\rm tot}^{\rm gal}(k,u,z)=P_{\rm clus}^{\rm gal}(k,u,z)+1/n_{\rm gal}(z)+N_{\rm sys}$ is the  total redshift-space galaxy power spectrum, where $N_{\rm sys}$ is the systematic noise. The detailed calculation of $P_{\rm tot}^{\rm gal}$ can be found in \cite{Gong19}. $N^{\rm cross}_{\rm m}$ is the number of modes for the cross power spectrum, which can be obtained by counting the $k$ modes in each wavenumber interval, considering the survey designs of both SPHEXEx and CSST.

In Figure~\ref{fig:Pl_cross}, we show the multipole moments of cross power spectra of H$\alpha$, [OIII] and [OII] and CSST spectroscopic galaxy survey at $z=1$ in the left, middle and right panels, respectively. We can see that since there is no interlopers appearing in the cross power spectrum, unlike the auto line power spectrum shown in Figure~\ref{fig:Pl_err}, the cross power spectra can reflect the strengths of the signal lines H$\alpha$, [OIII] and [OII] at $z=1$. The measurements of the cross power spectrum has relatively high SNR, which can obtained by replacing $P_{\ell,\rm obs}$ and $\Delta P_{\ell,\rm obs}$ by $P_{\ell}^{\rm cross}$ and $\Delta P_{\ell}^{\rm cross}$ in Eq.~(\ref{eq:SNR}). We find that, for $P_0$, $P_2$ and $P_4$, SNR=35.3, 3.7 and 1.8 for H$\alpha\times \rm gal$, 26.6, 2.7 and 1.4 for $\rm [OIII]\times gal$, and 28.0, 2.9 and 1.4 for $\rm [OII]\times gal$, respectively. We can find that H$\alpha\times \rm gal$ power spectrum has the highest amplitude with largest SNR, while $\rm [OIII]\times gal$ and $\rm [OII]\times gal$ are lower and have similar detectability. 

In addition to cross-correlating with traditional galaxy survey, we can also calculate the cross-correlations between emission lines, such as the cross-correlations of H$\alpha$, [OIII] and [OII]  at the same redshift between different frequency channels, or cross-correlation with 21-cm line measured by radio telescopes, e.g. Square Kilometer Array (SKA), Canadian Hydrogen Intensity Mapping Experiment (CHIME), and $\it Tianlai$ project \citep{Lidz16, Gong17}. This kind of cross-correlation is also helpful to reduce instrumental noise and contaminations of interloper lines, and hence can improve the constraint results. In this study, the cross-correlation with galaxy survey is sufficient and probably the best choice for discussion, and we will discuss other cross-correlations in details in the future work.

\section{model fitting}

\begin{figure*}
\centerline{
\includegraphics[scale=0.27]{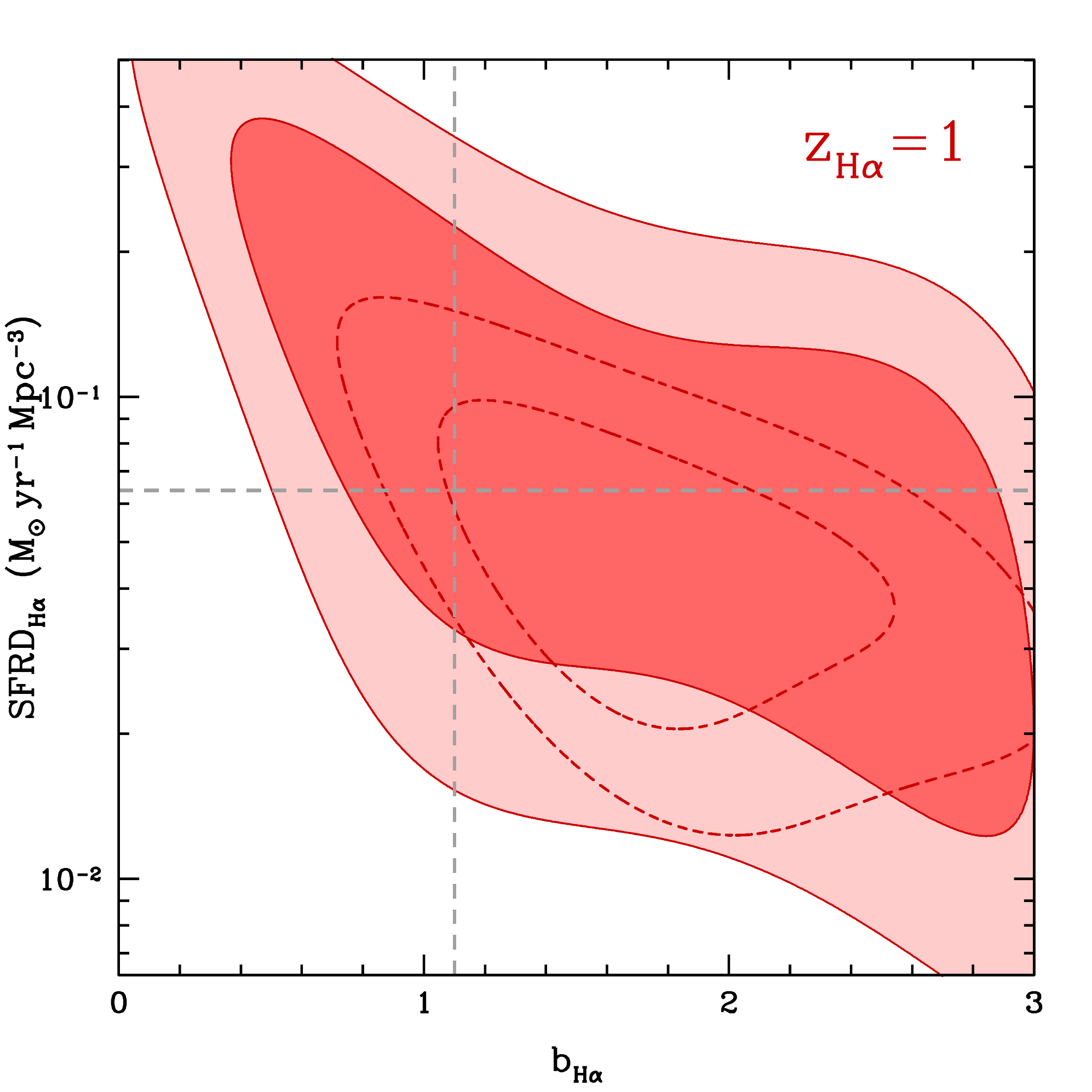}
\includegraphics[scale=0.27]{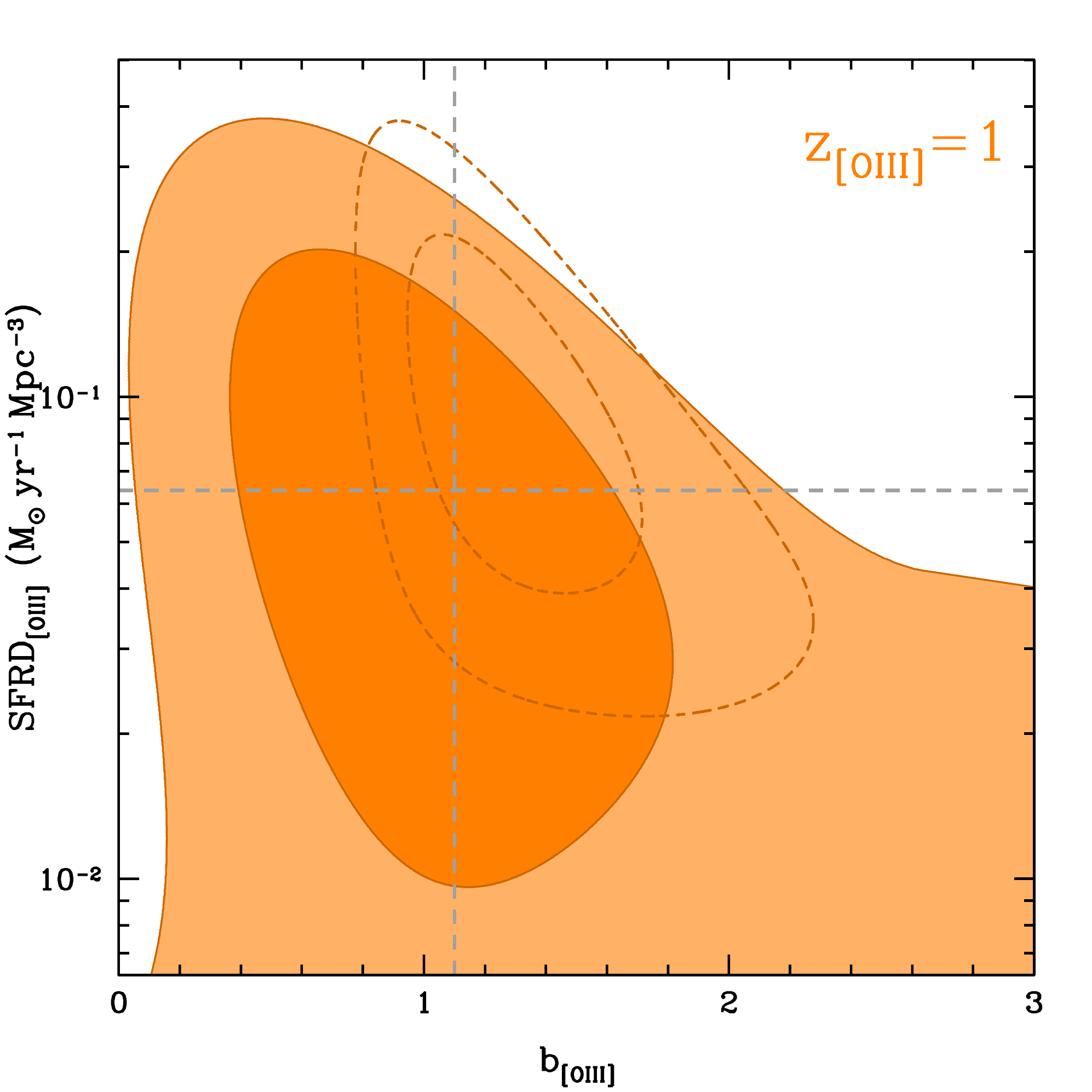}
\includegraphics[scale=0.27]{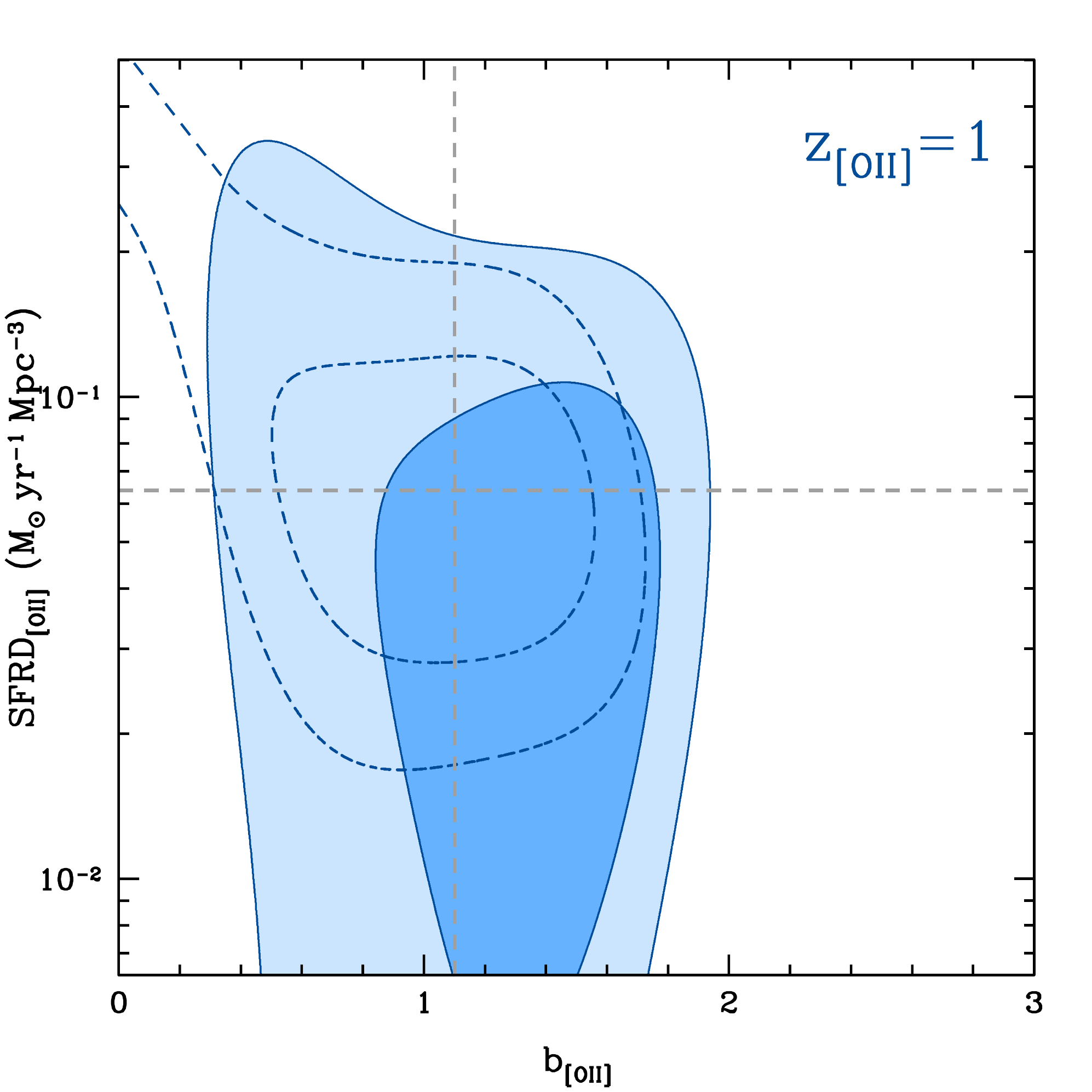}}
\caption{The contour maps of SFRD vs. $\bar{b}_{\rm line}$ of signal lines at $z=1$. The left, middle and right panels show the constraint results from the auto power spectra only (solid), and cross power spectra included (dashed) for H$\alpha$, [OIII] and [OII] lines, respectively. The gray dashed lines indicate the fiducial values of SFRD and line bias at $z=1$.}
\label{fig:b_SFRD_lines}
\end{figure*}

\begin{figure*}[t]
\centerline{
\includegraphics[scale=0.34]{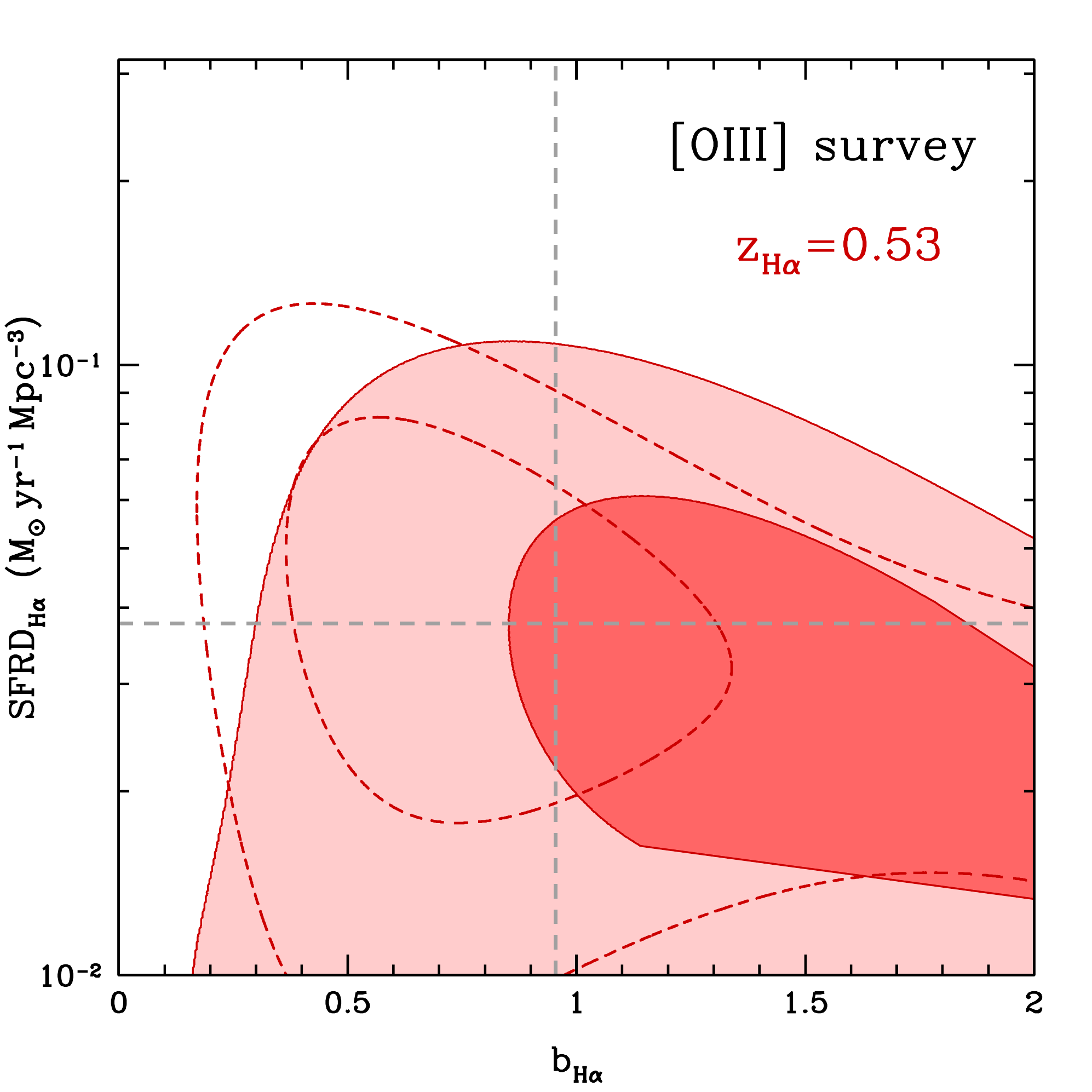}
\includegraphics[scale=0.34]{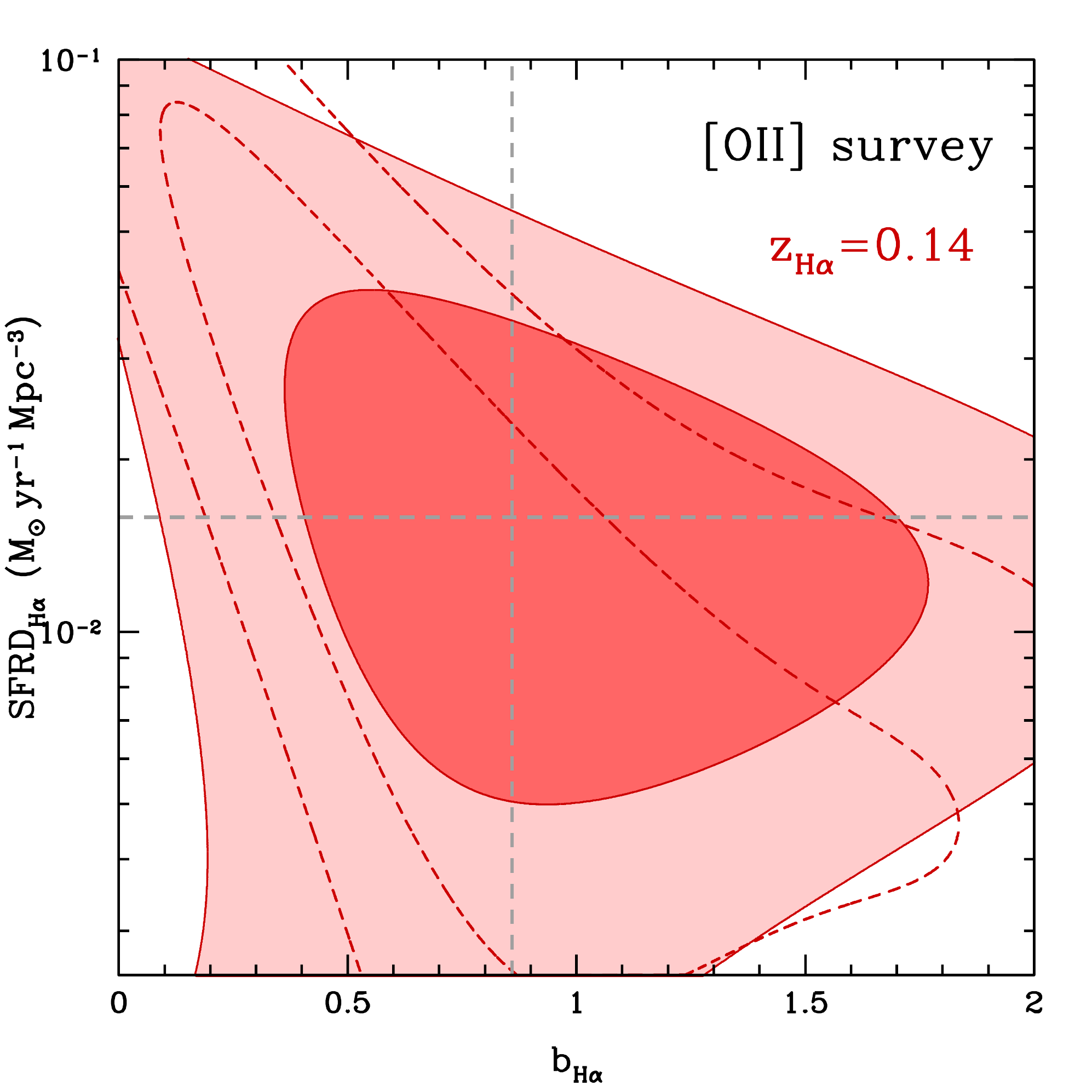}}
\centerline{
\includegraphics[scale=0.34]{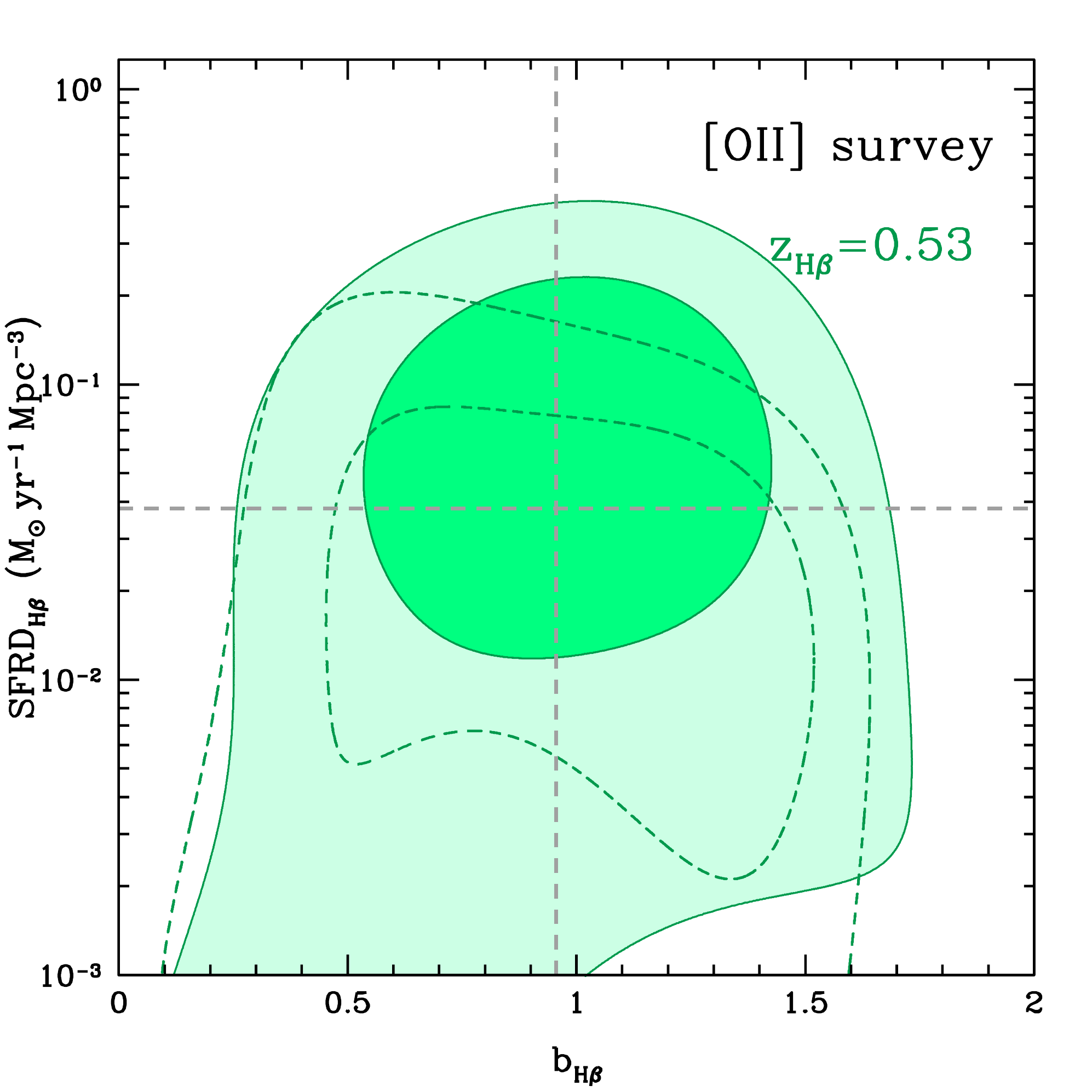}
\includegraphics[scale=0.34]{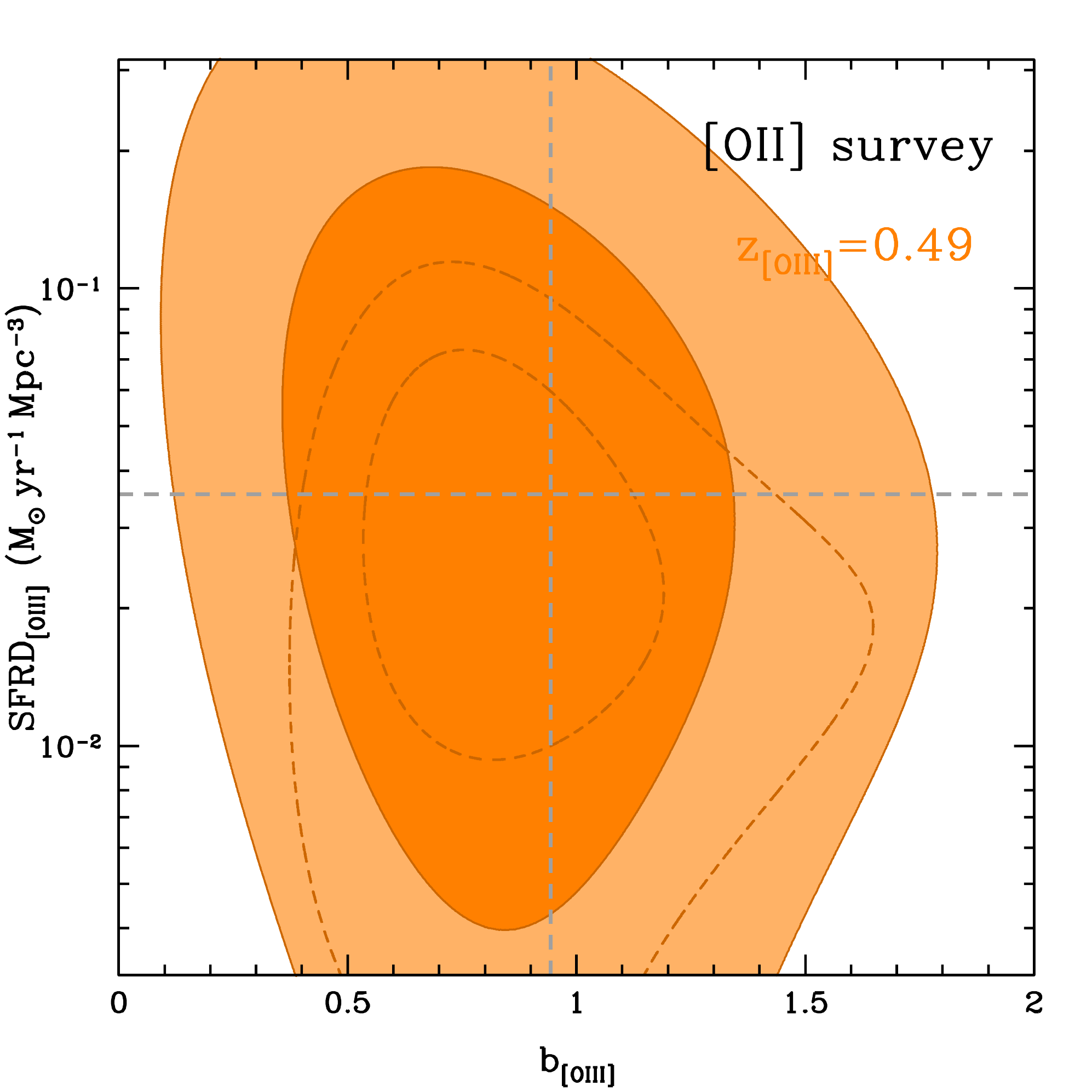}}
\caption{The contour maps of SFRD vs. $\bar{b}_{\rm line}$ at the redshifts of interloper lines for H$\alpha$, [OIII] and [OII] intensity mapping at $z=1$. The solid and dashed contours are the constraint results from auto only and cross power spectra included, respectively. In the top-left panel, the result of the interloper line H$\alpha$ at $z=0.53$ for [OIII] survey at $z=1$ is shown. In the top-right, bottom-left and -right panels, the results of the interloper lines H$\alpha$ at $z=0.14$, H$\beta$ at $z=0.53$ and [OIII] at $z=0.49$ for [OII] survey at $z=1$ are shown, respectively.}
\label{fig:b_SFRD_proj}
\end{figure*}

\begin{figure}[t]
\includegraphics[scale = 0.42]{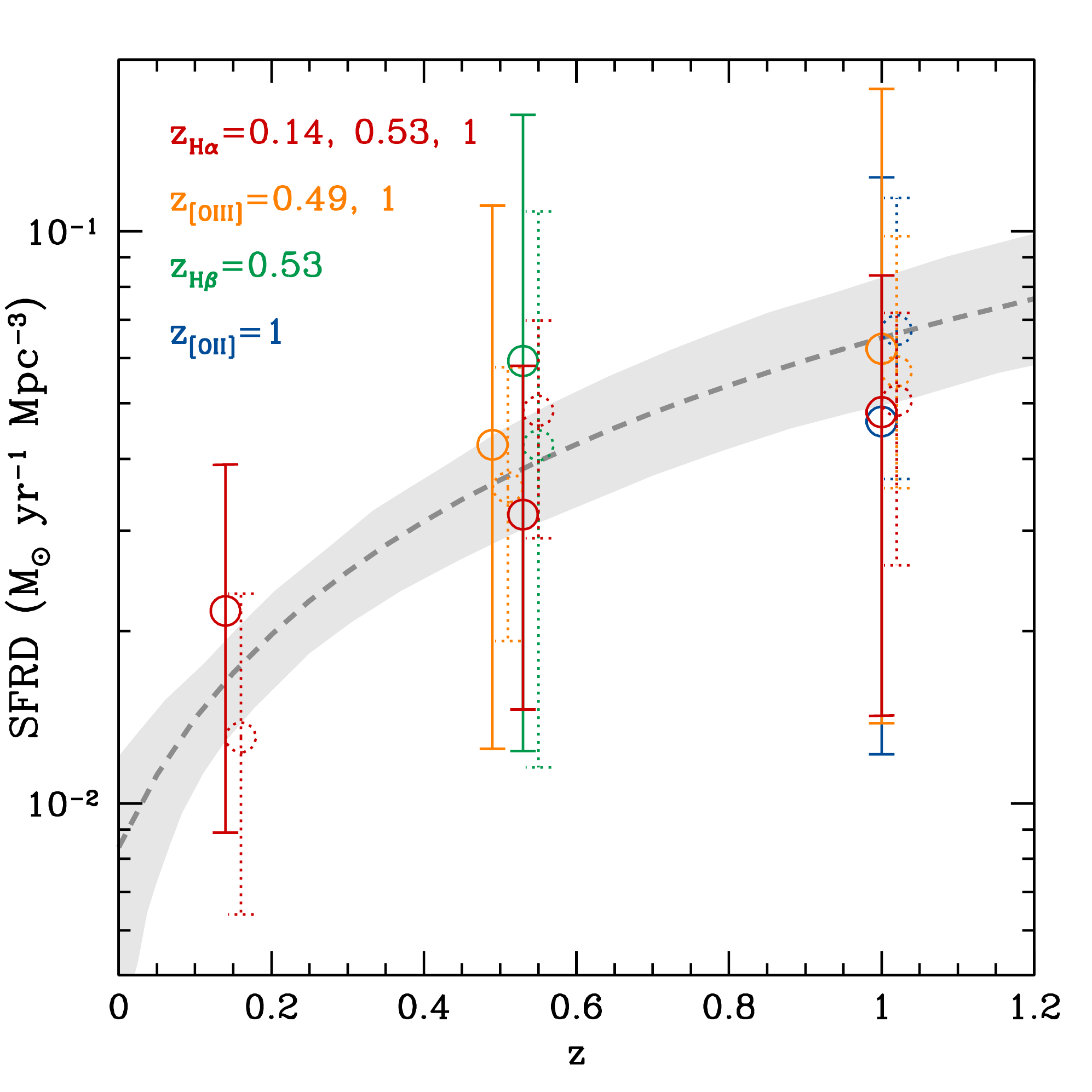}
\caption{The SFRD vs. $z$ derived from the constraint results by the H$\alpha$, [OIII] and [OII] intensity mapping survey. The gray dashed curve shows the fitting result given in \cite{Hopkins06}, and the shaded band denotes 2$\sigma$ C.L. The solid and dotted data points and error bars are derived from the fitting results of auto only and cross power spectra included cases, respectively. The dotted data points are shifted by +0.02 on redshift to show them clearly.}
\label{fig:SFRD_z}
\end{figure}

After generating mock data of multipole auto and cross power spectra for H$\alpha$, [OIII] and [OII] lines, we make use of the MCMC method to explore the constraints on the parameters of cosmological and astrophysical models. We have nine cosmological parameters in the model, and their fiducial values are $\Omega_{\rm b}=0.05$, $\Omega_{\rm M}=0.3$, $\sigma_8=0.8$, $n_{\rm s}=0.96$, $h=0.7$, $w_0=-1$, $w_a=0$, $\sum m_{\nu}=0$, $f_{\rm NL}=0$. Besides, we also set SFRD, mean line bias $\bar{b}_{\rm line}$ and total shot-noise power spectrum $P_{\rm shot}^{\rm tot}$ as free parameters for both signal and interloper lines. When cross power spectrum is involved in the constraints, the galaxy bias $b_g$ and shot-noise $P^{\rm shot}_{\rm cross}$ are included as free parameters. Hence we totally have 12 (9 cosmological parameters + $\rm SFRD_{H\alpha}$ + $b_{\rm H\alpha}$ + $P^{\rm shot}_{\rm H\alpha}$) and 14 (+ $b_{g}$ + $P_{\rm cross}^{\rm shot}$) free parameters for H$\alpha$ auto and cross power spectra, 14 (9 cosmological parameters + $\rm SFRD_{[OIII]}$ + $b_{\rm [OIII]}$ + $\rm SFRD^{\rm int}_{H\alpha}$ + $b^{\rm int}_{\rm H\alpha}$ + $P^{\rm shot}_{\rm tot}$) and 16 (+ $b_{g}$ + $P_{\rm cross}^{\rm shot}$) for [OIII], and 18 (9 cosmological parameters + $\rm SFRD_{[OII]}$ + $b_{\rm [OII]}$ + $\rm SFRD^{\rm int}_{H\alpha}$ + $b^{\rm int}_{\rm H\alpha}$ + $\rm SFRD^{\rm int}_{[OIII]}$ + $b^{\rm int}_{\rm [OIII]}$ + $\rm SFRD^{\rm int}_{H\beta}$ + $b^{\rm int}_{\rm H\beta}$ + $P^{\rm shot}_{\rm tot}$) and 20 (+ $b_{g}$ + $P_{\rm cross}^{\rm shot}$) for [OII], respectively.

We adopt $\chi^2$ method to perform the fitting process, and the $\chi^2$ for a multipole power spectrum is given by
\be
\chi^2 = \sum_{k\,{\rm bin}} \left[P_{i}^{\rm th}(k)-P_{i}^{\rm obs}(k)\right]{\rm Cov}_{ij}^{-1}\left[P_{j}^{\rm th}(k)-P_{j}^{\rm obs}(k)\right],
\ee
where $P_{i}^{\rm th}$ and $P_{i}^{\rm obs}$ are the theoretical and observed multipole power spectra with $\ell=0$, 2, and 4, respectively. ${\rm Cov}_{ij}$ is the covariance of the power spectrum. The chi-square of line auto power spectrum $\chi^2_{\rm auto}$ can be obtained by using Eq.~(\ref{eq:P_obs}) and Eq.~(\ref{eq:Del_Pk}), and Eq.~(\ref{eq:P_cross}) and Eq.~(\ref{eq:Del_Pcross}) for calculating  $\chi^2_{\rm cross}$. Then the total chi-square is given by $\chi^2_{\rm tot}=\chi^2_{\rm auto}+\chi^2_{\rm cross}$. The likelihod function can be calculated by $\mathcal{L}\sim {\rm exp}(-\chi^2/2)$.

In the MCMC technique we adopt, the Metropolis-Hastings algorithm is used to determine the accepted probability of a new chain point \citep{Metropolis53,Hastings70}. The proposal density matrix is obtained from a Gaussian sampler with adaptive step size \citep{Doran04}. The flat priors are assumed for all the free parameters with large parameter ranges. We run 20 parallel chains for each case we study, and get about 100,000 chain points for each chain after convergence criterion is fulfilled \citep{Gelman92}. After performing the burn-in and thinning processes for each chain, we combine all chains together. Finally, we obtain about 10,000 chain points for ploting 1D and 2D probability distribution functions (PDFs) of the free paramters.

\section{constraint results}
\label{sec:results}

In Figure~\ref{fig:m0s8_w0wa_lines}, we show the contour maps of $\Omega_{\rm M}$ vs. $\sigma_8$ and $w_0$ vs. $w_a$ in the upper and lower panels for H$\alpha$, [OIII] and [OII] observations, respectively. The constraint results from line auto power spectrum only that measured by SPHEREx experiment are shown in solid contours, and the dashed contours denote the results when including cross power spectrum detected by CSST galaxy survey. As can be seen, the constraint results are consistent with the fiducial values of the parameters (in gray parallel and vertical lines) in 1$\sigma$ confidence level (C.L.).We can find that the constraint results of [OIII] and [OII] lines are basically better ($w_0$ vs. $w_a$) than or comparable ($\Omega_{\rm m}$ vs. $\sigma_8$) to H$\alpha$, although interlopers appear in the two former lines. After including the cross power spectrum with CSST galaxy survey, the constraints can be evidently further improved as shown in dashed contours.

In Figure~\ref{fig:mv_fnl_lines}, the 1D PDFs of $\sum m_{\nu}$ and $f_{\rm NL}$ are shown. The solid and dashed curves are for line auto only and cross power spectrum included cases, respectively. We find that the total neutrino mass can be constrained as $\sum m_{\nu}\lesssim0.3$ eV at 1$\sigma$ C.L. for all the three emission lines, and the results can be more stringent when including the cross power spectra. For the primordial non-Gaussianity parameter, we find that $|f_{\rm NL}|\lesssim15$ for H$\alpha$ line, and $|f_{\rm NL}|\lesssim10$ for [OIII] and [OII] lines. When considering cross power spectrum with CSST galaxy survey, $f_{\rm NL}$ can be further constrained as $|f_{\rm NL}|\lesssim7$ for [OIII] and [OII] lines, $|f_{\rm NL}|\lesssim11$ for H$\alpha$ case. As the same as the constraints on $\Omega_{\rm M}$ vs. $\sigma_8$ and $w_0$ vs. $w_a$ shown in Figure~{\ref{fig:m0s8_w0wa_lines}}, [OIII] and [OII] line intensity mapping can provide good constraints on neutrino mass and primordial non-Gaussianity in our method, which are even better than that from H${\alpha}$ line without contamination of interloper lines. This indicates that if we have good understanding on the interloper lines, they are actually can be seen as signals as well, and can help to constrain the cosmological parameters at different epochs of the Universe.

In Figure~\ref{fig:b_SFRD_lines} and Figure~\ref{fig:b_SFRD_proj}, we show the contour maps of SFRD vs. $\bar{b}_{\rm line}$ for H$\alpha$, [OIII] and [OII] at $z=1$ and that for the interloper lines at corresponding redshifts, respectively. We find that the constraint results are consistent with the fiducial values of SFRD and $\bar{b}_{\rm line}$ in 1$\sigma$ C.L. After adding the cross power spectra in the fitting process, we can get apparently better constraint results. We also find that the contours of  SFRD vs. $\bar{b}_{\rm line}$ are not regular, that even the degeneracy direction is not quite obvious in some case. This is because that the ``nuisance'' parameters, such as SFRD and line bias parameters of the other (signal or interloper) lines and shot-noise terms, can significantly disturb the shape of parameter space and result in irregular parameter contours.

In addition, we can find that the perfect degeneracy between SFRD (or mean intensity $\bar{I}_{\rm line}$, see Eq.~(\ref{eq:I_SFR})-(\ref{eq:SFRD_z}) for details) and line bias $\bar{b}_{\rm line}$ can be broken to some extent by adopting multipole moments of power spectrum with redshift distortion as shown in both Figure~\ref{fig:b_SFRD_lines} and Figure~\ref{fig:b_SFRD_proj}. This provides a support for the discussion in \S\ref{subsec:sps} about the advantage of using redshift-space power spectrum \citep{Lidz16,Chen16}. Besides, since we can obtain good constraints on SFRD and $\bar{b}_{\rm line}$ for both signal and interloper lines, it implies that the power spectra of the signal and interloper lines can be distinguished in the fitting process as we discuss and show in \S\ref{subsec:ops} and Figure~\ref{fig:ratio_lines}. This proves that the multipole moments of redshift-space power spectrum is feasible and effective for extracting the information of the cosmological and astrophysical quantities. 

We show the best-fitting SFRD and 1$\sigma$ error for the signal and interloper lines as a function of redshift in Figure~\ref{fig:SFRD_z}. The gray dashed curve denotes the fiducial values of SFRD at different redshifts given by \cite{Hopkins06}. We can see the fitting results are consistent with the fiducial values in 1$\sigma$ C.L. We notice that our results is worse than the prediction given by \cite{Gong17}. This is because it is actually an optimistic estimate in \cite{Gong17}, that only SFRD is set as free parameter and Fisher matrix method is simply adopted in that work. In this study, we have included much more number of components and free parameters in the model, and use MCMC technique with mock data to obtain a more realistic and reliable results.

We should also note that the constraint results shown in this section are only derived from the observations at $z=1$ as an example. At $z>1$, the constraints may not as strong as that at z=1 considering strength of signal line, interloper lines, and cross-correlations with ordinary galaxy surveys. However, in a real survey like SPHEREx, more lower and higher redshifts of the signal lines will be explored at different bands in the mean time, and a lot more data can be obtained and used to constrain the cosmological and astrophysical parameters simultaneously. This will undoubtedly provide tighter constraints on these parameters, and can be even better than ordinary galaxy survey, especially for the Universe at high redshifts.

\section{Summary}
\label{sec:summary}

In this work, we propose to use the multipole moments of redshift-space intensity power spectrum of emission line for constraining the cosmological and astrophysical parameters. In principle, the multipole power spectrum can effectively distinguish the signal and interloper emission lines, and break degeneracy between the line mean intensity and bias in the model.

We include time-variable equation of state of dark energy, massive neutrinos, and primordial non-Gaussianity in the model, which can change both kinematical and dynamical evolution of the Universe. The mean line intensity is estimated by the SFRD derived from observations. Then we calculate the multipole moments of redshift-space intensity power spectra of H$\alpha\,6563\rm \AA$, [OIII]$\,5007\rm\AA$ and [OII]$\,3727\rm \AA$ at $1\le z\le3$. In order to discuss observed power spectra with interloper lines, we evaluate the total observed multipole power spectra of the three emission lines at $z=1$, and the uncertainties of the power spectra from observations are also estimated. 

In the discussion of line detection, we take SPHEREx experiment as an example to explore the measurements of the three emission lines, and also compute cross power spectra by including the CSST spectroscopic galaxy survey. The MCMC method is adopted for fitting the cosmological and astrophysical parameters in the model for the three emission lines with interlopers.

We find that the cosmological and astrophysical parameters can be properly constrained, and the best-fits are consistent with the fiducial values in 1$\sigma$ C.L. This provides strong supports for the advantages of multipole power spectrum for extracting information of cosmological and astrophysical quantities, and proves that this method is feasible and effective. The constraints can be further improved by involving the measurements from other redshifts and cross-correlations between different lines, that can even get better results than traditional galaxy surveys, especially for probing the Universe at high redshifts.

\begin{acknowledgments}
YG acknowledges the support of NSFC-11773031, NSFC-11822305, NSFC-11633004, MOST-2018YFE0120800, the Chinese Academy of Sciences (CAS) Strategic Priority Research Program XDA15020200, the NSFC-ISF joint research program No. 11761141012, and CAS Interdisciplinary Innovation Team.
\end{acknowledgments}


\end{document}